\documentclass[ 
	aps,prd,
	oneside, 
	showpacs,
	floatfix,
	nofootinbib,
	notitlepage
]{revtex4-1}  
\usepackage{verbatim}
\usepackage[  
        hmargin=3.7cm %to make horizontal margins equal to 3cm.
]
{geometry}

\usepackage{graphicx}
\usepackage{amssymb}
\usepackage{amsmath}
\usepackage{tensor} %Tensoren
\usepackage{amsfonts}
\usepackage{bbold}
\usepackage{xcolor}
\usepackage[pdfstartview=FitBV,bookmarksnumbered,
bookmarksopen,
colorlinks =true,
, linkbordercolor=blue
 ,linkcolor = black
 ,
citecolor = blue,
urlcolor = blue
]{hyperref}
\makeatletter
\renewcommand\frontmatter@abstractwidth{\dimexpr\textwidth-1in\relax}
\makeatother

\begin{document}
\title{Non-perfect-fluid space-times in thermodynamic equilibrium
 %\texorpdfstring{\\}{}
and generalized Friedmann equations}

\author{Konrad Schatz}
\email{konscha@physik.tu-berlin.de}
\affiliation{Institut f\"{u}r Theoretische Physik, Technische Universit\"{a}t
Berlin,\\
Hardenbergstra\ss e 36, D-10623 Berlin, Germany }

\author{Horst-Heino von Borzeszkowski}
\email{borzeszk@mailbox.tu-berlin.de}
\affiliation{Institut f\"{u}r Theoretische Physik, Technische Universit\"{a}t
Berlin,\\
Hardenbergstra\ss e 36, D-10623 Berlin, Germany }
  
\author{Thoralf Chrobok}
\email{tchrobok@mailbox.tu-berlin.de}
\affiliation{Institut f\"{u}r Theoretische Physik, Technische Universit\"{a}t
Berlin,\\ 
Hardenbergstra\ss e 36, D-10623 Berlin, Germany }

\begin{abstract}
We determine the energy-momentum tensor of non-perfect fluids in thermodynamic
equilibrium. 
To this end, we derive the constitutive
equations for energy density,
isotropic and anisotropic pressure as well as for heat-flux from the
corresponding propagation equations and by drawing on Einstein's equations.
Following Obukhov at this, we assume the corresponding space-times to be
conform-stationary and homogeneous.
This procedure provides these quantities in closed form, i.e.,
 in terms of the structure constants of the three-dimensional
isometry group of homogeneity and, respectively, in terms of the kinematical
quantities expansion, rotation and acceleration.
In particular, we find a generalized form of the Friedmann equations.
As special cases we recover Friedmann and
G\"odel models as well as non-tilted Bianchi solutions with anisotropic pressure. 
All of our results are derived without assuming
any equations of state or other specific thermodynamic conditions a priori.
For the considered models, results in literature
are generalized to rotating fluids with dissipative fluxes.
\end{abstract}


\maketitle

\section{Introduction}

In this paper, we consider systems described by Einstein's equations
\begin{align} %\notag
\label{eq:einstein-field-eq}
R_{ab}-\frac{1}{2} R g_{ab}= T_{ab}
\end{align}
with an energy-momentum
tensor and equations of state neither of which
 are specified by any \textit{ad hoc} assumptions. Instead,
we discuss the whole question from a thermodynamic
perspective\footnote{We emphasize that we approach this
without any specific thermodynamic conditions as done for instance in
\cite{ellis1969rel-cosm}, i. e.,
we refrain from applying linear or extended thermodynamics.}.
This consideration is discussed for a class of cosmologically interesting
metrics (introduced and shown to be observationally admissible in \cite{chrobok2002, obukhov1998, obukhov1990bianchi_list, obukhov2000}, see (\ref{eq:metric_allg_obuk}) below). In terms of the temperature and the
kinematic invariants characterizing the matter, our consideration provides a class of general equations of state
(``matter equations'') which are compatible with Einstein's equations and correspond to
generalized Friedmann equations. This framework can find (and has found)
applications in relativistic cosmology and astrophysics. Basically, it allows to
go beyond the standard phase cosmology (governed by phases with certain equations of state like inflation,
radiation and dust that have to be fitted by fine-tuning) and to describe the cosmological state transitions 
from phase to phase by intermediate stages. 
 However, in \cite{NOJIRI1, NOJIRI2, CAPOZIELLO2006, SEBASTIANI2014} a fine
 tuned sudden passage from the decelerated to the accelerated regime, as observed today, 
 produces inadequacies. These are then avoided by {\it ad hoc} introduced
 equations of state where viscosity originates from geometry (e.g. $H,
 \dot{H}$).
 Our calculations can provide a theoretical
foundation of such equations. Furthermore, this framework also contributes to a
physical discussion of no-go theorems like the shear-free fluid conjecture
\cite{ellis1971}.
For instance, this thermodynamic
approach enables one to sharpen the theorem (proved in \cite{borzchrobunknown},
without explicitly referring to thermodynamics)
which states for non-vanishing acceleration that rotation
and expansion cannot
simultaneously be equal to zero: in \cite{borzchrobunknown}
it has been shown that models with vanishing acceleration
do not allow for non-vanishing rotation.

  Ehlers, Geren and Sachs have proven \cite{EGS} that the high isotropy of
  the cosmic microwave background (CMB) and the vanishing of 
  shear\footnote{For the definition of the quantities: shear, rotation,
  acceleration and expansion see \cite{ellis1969rel-cosm} or see also Sec.
  \ref{sec:tetrads-kin-invariants}.} $\sigma_{ab}$ of a congruence of curves are closely related 
  to the requirement that there exists a conformal Killing vector
  field\footnote{In \cite{Duggal}, it is shown that, while for $\sigma_{ab}=0$ fluids the CKV property is essential, for $\sigma_{ab}\neq 0$ fluids this property has to be generalized to conformal collineation.} 
  being parallel to the tangents of the curves (i.e., to the velocity of a streaming fluid).
  In detail, it was shown (Lem. 3 in \cite{EGS}) that a space-time admits a
  time-like conformal Killing vector field (CKV) \begin{align} %\notag
\label{eq:diffgeo_conform_killing}
\xi_{a;b}+\xi_{b;a}=\Phi g_{ab}
\end{align}
with $\xi_{a}=\alpha u_a$, if and only if there is a velocity
field $u_a$ with $u_a u^a=-1$ satisfying
\begin{align} \label{BEDINGUNG1}
\sigma_{ab}=0 \quad \text{and} \quad 
	\left(
		\dot{u}_{[a}-\frac{\Theta}{3}u_{[a}
	\right)\vphantom{}_{;b]}=0
\end{align}
where $\dot{u}_a$ is the acceleration and $\Theta$ the expansion. 
Olivier and Davis \cite{OlivierDavis} showed that (\ref{BEDINGUNG1}) 
is a necessary and sufficient condition for the existence of a CKV in the case
of rotating space-times, too. In the following, we consider such
conform-stationary space-times\footnote{According to \cite{hasse1988,coley1991} this is equivalent to parallax-free cosmological models.}.

In particular, the second condition allows to imply a parameter $\alpha$ 
which can be identified as the inverse temperature, $\alpha=1/T$, so that
$\xi_{a}=u_a/T$ can be interpreted as temperature vector. 
This parameter occurs if the second equation in (\ref{BEDINGUNG1}) is rewritten
as (Thm. 2.1 in \cite{OlivierDavis})
\begin{align} \label{BEDINGUNG}
\frac{\alpha_{,a}}{\alpha}=\dot{u}_a-\frac{\Theta}{3}u_a.
\end{align}

Additionally, we assume that the considered space-times are spatially
homogeneous.
This reduction to Bianchi-type models still allows for the matter
distribution to be anisotropic, while the CMB is isotropic.

Altogether, we are led to the subclass of tilted Bianchi models constructed by
Obukhov \cite{chrobok2002, obukhov1998, obukhov1990bianchi_list, obukhov2000} that admit a CKV, 
\begin{align} \notag \label{eq:metric_allg_obuk}
  ds^2 ={}& \tensor{g}{_a_b} \, dx^a \, dx^b \\
  	={}& - dt^2 + 2\, a\, n_{\hat{a}} \, dx^{\hat{a}} dt
  	+ a^2 \, \tensor{\gamma}{_{\hat{a}\hat{b}}} \, dx^{\hat{a}} dx^{\hat{b}} \,.
\end{align}
Thereby, rotating and expanding models with acceleration and
isotropic CMB are considered in \cite{chrobok2002, obukhov1998,
obukhov1990bianchi_list, obukhov2000}. In contrast to that, 
in \cite{CLARK1, CLARK2, CLARK3}, 
Ehlers-Geren-Sachs theorems were, partly in a
generalized version, used to study and determine a class of space-times
containing also inhomogeneous cosmological models, 
with non-trivial acceleration but zero rotation.

To complete notation used in \eqref{eq:metric_allg_obuk}, we define
\begin{align} \notag
			\tensor{n}{_{\hat{a}}} = \tensor{\nu}{_{\hat{\mu}}}\,
			\tensor{e}{^{\hat{\mu}}_{\hat{a}}}
\qquad
\text{and}
\qquad
		    \tensor{\gamma}{_{\hat{a} \hat{b}}} = \tensor{\beta}{_{\hat{\mu} \hat{\nu}}}\,
		    \tensor{e}{^{\hat{\mu}}_{\hat{a}}}\, \tensor{e}{^{\hat{\nu}}_{\hat{b}}}
\end{align}
with $\tensor{\nu}{_{\hat{\mu}}}$ and $\tensor{\beta}{_{\hat{\mu}
\hat{\nu}}}$ as arbitrary constants. Here the triad components
$\tensor{e}{^{\hat{\mu}}_{\hat{a}}} =
\tensor{e}{^{\hat{\mu}}_{\hat{a}}}(x^{\hat{k}})$
form a basis which is invariant under 
the spatial isometries of the
 Bianchi models. Accordingly, their Lie derivative with respect to
 the generating Killing vector fields (KV) vanishes
 (for details see \cite{stephani2003exact}).
The components
$\tensor{e}{^{\hat{\mu}}_{\hat{a}}}(x^{\hat{k}})$ 
are supposed to be functions of the space-like canonic coordinates only
and determine the metric \eqref{eq:metric_allg_obuk} as to the Bianchi-type.
The coordinate $t = x^0$ denotes the proper time with
respect to a fluid particle and $a = a(t)$ is the scale factor.

Furthermore, Latin indices are used for the coordinate
while Greek indices are used for the tetrad description of the tensor
components.
All indices with hat denote the three spatial dimensions, e.g.
$\hat{a} = {1,2,3}$ and $\hat{\alpha} = {(1),(2),(3)}$, whereas those without
hat run through all four space-time dimensions, $a = {0,1,2,3}$
and $\alpha={(0),(1),(2),(3)}$.

Regarding the thermodynamic proposition, we follow the Eckart approach and
assume that the model under consideration is in thermodynamic equilibrium. The Eckart approach to the relativistic Theory of Irreversible
Processes \cite{eckart1940} (see also \cite{israel1989fuid-intro,weinberg1972,neugebauer1980}) is based on the balance equations
for the particle number (where $\mu$ represents mass density and $v = (1/\mu)$
the specific volume)
\begin{align} \label{eq:gleichgewicht_teilchen-zahl-erhalt}
	\left( \mu \tensor{u}{^a} \right)_{;a} = 0 \, ,
\end{align}
the energy-momentum\footnote{In particular, regarding
\eqref{eq:gleichgewicht_teilchen-zahl-erhalt}, the null-component can be interpreted as the first law of thermodynamics \cite{borzchrob2011, schellstede2013},
 i.e., as the conservation of internal energy.}
\begin{align}\label{eq:thermo-equil_ei-tensor_erster_hpt-satz}
    \tensor{T}{^{a b}_{; b}} = 0\,
\end{align}
and the entropy 
\begin{align}
	\sigma = \tensor{s}{^a_{;a}} \geq 0 \, 
\end{align}
where $\tensor{s}{^a}$ denotes the entropy vector and $\sigma$ the density of
the non-negative entropy production.
 In the case of thermodynamic equilibrium
(vanishing of entropy production), appropriated supplementary conditions have to be added by hand.
In the general-relativistic
version of this theory the framework is completed by Einstein's
gravitational equations \eqref{eq:einstein-field-eq}.
However, in the general-relativistic Theory of Irreversible
Processes no further assumptions have to be introduced \textit{ad hoc}
in order to yield thermodynamic equilibrium \cite{borzchrob2006}.

Now, if the entropy vector is defined according to
\cite{israel1989fuid-intro,stephani1988,neugebauer1980},
\begin{align}
	\tensor{s}{^a} = \mu\, s\, \tensor{u}{^a} + \frac{\tensor{q}{^a}}{T} \, ,
\end{align}
the entropy production can be reformulated as
\begin{align} %\label{eq:gleichgewicht_entropie-prod_zw1} \notag
	\sigma
	=
	\frac{\tensor{u}{_b}}{T}
	\left(
		\tensor{T}{^{a b}}
		- \rho\, \tensor{u}{^a}\, \tensor{u}{^b}
		- p\, \tensor{h}{^{a b}}
	\right)_{; a}
	+
	\left(\frac{\tensor{q}{^a}}{T}\,  \right)_{;a} \, .
\end{align}
Here $q^{a}$ denotes the heat-flux, $\rho$ the
energy density, $p$ the pressure, $u^a/T$ the temperature vector and
$\tensor{h}{^{a b}} = \tensor{g}{^{a b}} + \tensor{u}{^a}\, \tensor{u}{^b}$.
Finally, by decomposing the energy-momentum tensor
\eqref{eq:thermo-equil_ei-tensor_zerleg}
this yields \cite{stephani1988}:
\begin{align} \label{eq:gleichgewicht_entropie-prod_bed-end}
	\sigma
	=
	-
	\left( \frac{\tensor{u}{_b}}{T} \right)_{; a}
	\left(
		\tensor{T}{^{a b}}
		- \rho\, \tensor{u}{^a}\, \tensor{u}{^b}
		- p\, \tensor{h}{^{a b}}
	\right)  \, .
\end{align}
As shown in \cite{borzchrob2006}, regarding
the conformal Killing equation \eqref{eq:diffgeo_conform_killing}, 
the second term in brackets turns out to be
 traceless which results in a vanishing entropy production
$\sigma$:\footnote{This shows that non-perfect fluids are not necessarily
incompatible with reversible thermodynamics
\cite{stephani1988,bedran1993,triginer1995}. However, for space-times without a
CKV or KV \eqref{eq:entropy-prod-cond} can only be solved by assuming a perfect
fluid.}
\begin{align} \label{eq:entropy-prod-cond}
	\left(
		\tensor{T}{^{a b}}
		-
		\rho\, \tensor{u}{^a}\, \tensor{u}{^b}
		-
		p\, \tensor{h}{^{a b}}
	\right)
	\left(
		\tensor{\xi}{_{a; b}} + \tensor{\xi}{_{b; a}}
	\right)
	=
	0 \, .
\end{align}
That this CKV property is not purely mathematical, but has also a physical
meaning, is supported by the following arguments:

Firstly, the derivations of (\ref{eq:gleichgewicht_entropie-prod_bed-end}) 
and (\ref{eq:entropy-prod-cond}), given in \cite{stephani1988}, show that the
quantities $\rho$, $p$ and $T$ are thermodynamically well-determined. 
Indeed, it is assumed that the specific entropy $s$ 
is given as a function of the specific internal energy $u$ and the specific volume $v$, i. e.
\begin{align} \label{eq:entropy-dependencies}
	s = s(u, v) \, ,
\end{align}
so that (Gibbs equation\footnote{Moreover, for a co-moving observer the relation
$\rho = \mu (1 + u)$ holds.})
\begin{align} \label{eq:gleichgewicht_gibbs-gl}
		T\, \mathrm{d}s = \mathrm{d}u + p\, \mathrm{d}v		
\end{align}
and
\begin{align} \label{eq:entropy-derivs}
	\frac{\partial s}{\partial u}
	=
	\frac{1}{T}
	\quad , \quad
	\frac{\partial s}{\partial v}
	=
	\frac{p}{T} \, .
\end{align}
For the thermodynamic quantities defined in this way
\eqref{eq:entropy-prod-cond} is valid. Secondly, \eqref{eq:entropy-prod-cond} has the following solutions: Either the
fluid is perfect or the temperature vector has to be a CKV (containing
the special case of a Killing vector field). Therefore, the CKV property is
justified by defining equilibrium states.

This is confirmed by the fact that $\xi^a = u^a/T$, being a CKV, leads to 
some well-known models like Friedmann's and G\"odel's space-times with the
corresponding equations of state (see Sec. \ref{sec:matter-eqs_special-cases}).
Furthermore, it should be emphasized that the justification of the
thermodynamic meaning of the CKV condition given via
\eqref{eq:entropy-prod-cond}, i.e., in the context of phenomenological continuum theory, is supported by
considerations in the framework of kinetic theory, where the CKV
property of $\xi^a$ in combination with related equations of state for some
special cases is derived from Boltzmann's equation \cite{ZIMDAHL1, ZIMDAHL2, ZIMDAHL3, ZIMDAHL4}.

Based on the existence of such a CKV one can derive
a set of four propagation equations for non-perfect fluids
(see \cite{borzchrobunknown}), which link the propagation of
the matter content to the kinematic description of the space-time (see Sec.
\ref{sec:matter-eqs}).

The paper is organized as follows: In Sec. \ref{sec:tetrads-kin-invariants},
we introduce a suitable tetrad frame that allows us to establish
manageable equations. In addition, the decomposition of
the energy-momentum tensor with respect to
kinematic invariants is shortly reviewed and their form in tetrads for the
space-times \eqref{eq:metric_allg_obuk} is derived. 
Subsequently, by solving the propagation equations, we deduce in Sec.
\ref{sec:matter-eqs} general expressions for the whole matter content depending on the
structure constants and the kinematic invariants, respectively.
 After checking the consistency with Einstein's equations the general case of a
non-perfect fluid as well as particular cases like non-tilted \cite{kingellis1973} and stationary models are discussed.
 Among the special cases that can be recovered are
 the Friedmann and the G\"odel models.
 %Details about the calculations can be found in \cite{schatz2012dipl}.
 In Sec. \ref{sec:Interpretation} we discuss the results and provide alternative
 formulations, relevant for further observational and thermodynamic
 considerations.
\section{Tetrad formulation and kinematic invariants}
\label{sec:tetrads-kin-invariants}
In the following, we introduce
tetrads (see, e.g., \cite{chandra1983}) that allow for a convenient separation of
the variable objects,
$a(t)$ and $\tensor{e}{^{\hat{\mu}}_{\hat{a}}}(x^{\hat{k}})$,
 and the constants,
 $\tensor{\nu}{_{\hat{\mu}}}$ and $\tensor{\beta}{_{\hat{\mu} \hat{\nu}}}$ in
 \eqref{eq:metric_allg_obuk}.
Defining
\begin{align} \label{eq:metric_triad_embed}
	\tensor{\hat{e}}{^{\alpha}_{b}}
	:=
 \begin{pmatrix}
	  0 & 0 \\
	  0 & \tensor{e}{^{\hat{\alpha}}_{\hat{b}}} \\
 \end{pmatrix} \; {\text{ and }} \;
 \tensor{\check{e}}{_{\alpha}^{b}}
	:=
 \begin{pmatrix}
	  0 & 0 \\
	  0 & \tensor{e}{_{\hat{\alpha}}^{\hat{b}}} \\
 \end{pmatrix} \,
\end{align}
with
\begin{align} \label{eq:metric_kronecker_raumartig_embed}
  \tensor{\check{e}}{_{\alpha}^{b}}\, \tensor{\hat{e}}{^{\beta}_{b}}
  =
 \begin{pmatrix}
	  0 & 0 \\
	  0 &  \tensor*{\delta}{*_{\hat{\alpha}}^{\hat{\beta}}}\\
 \end{pmatrix} \, ,
\end{align}
the tetrads can be chosen as
\begin{align} \label{eq:metric_alter_tetrad}
	\tensor{\theta}{^{\alpha}_{b}}
	=
 	\tensor*{\delta}{*^{\alpha}_{0}} \, \tensor*{\delta}{*^{0}_{b}} + a\, \tensor{\hat{e}}{^{\alpha}_{b}} \; {\text{ and }} \;
 \tensor{\theta}{_{\alpha}^{b}}
	=
 	\tensor*{\delta}{*_{\alpha}^{0}} \, \tensor*{\delta}{*_{0}^{b}} + a^{-1}\, \tensor{\check{e}}{_{\alpha}^{b}} \, .
\end{align}
To fulfill the relations
\begin{align} \label{eq:tetrad_def}
	\tensor{g}{_a_b} ={}& \tensor{\zeta}{_\mu_\nu} \,			
		\tensor{\theta}{^\mu_a} \, \tensor{\theta}{^\nu_b} \, ,
\end{align}
the constant and symmetric matrix
$\tensor{\zeta}{_\mu_\nu}$ has to take the form
\begin{align} \label{eq:metric_tetmet}
		\tensor{\zeta}{_\mu_\nu} =
			\begin{pmatrix}
			  -1 & \nu_{\hat{\nu}}  \\
			  \nu_{\hat{\mu}} & \beta_{\hat{\mu} \hat{\nu}} \\
      \end{pmatrix} \, .
\end{align}
The structure constants of the isometry groups acting on the space-like
hypersurfaces and specific to the Bianchi models
can
be expressed by a 4-dimensional representation,
\begin{align} \label{eq:metric_struct-const_embed}
	\tensor{\hat{C}}{^\gamma_{\beta \alpha}}\,
	:=
		2\, \tensor{\check{e}}{_{[{\alpha}}^{b}_|}\, \tensor{\hat{e}}{^{\gamma}_{b,c|}}\,
		\tensor{\check{e}}{_{\beta]}^{c}} \, ,
\end{align}
such that
$\tensor{\hat{C}}{^\gamma_{0 \alpha}} = 0$, $\tensor{\hat{C}}{^\gamma_{\beta 0}} = 0$ and
$\tensor{\hat{C}}{^0_{\beta \alpha}} = 0$.
Expressions for the curvature tensors and scalars in terms of these newly
introduced tetrads are derived in Appendix \ref{sec:appendix-curv}.

On the basis of these preliminaries, we now introduce the kinematic invariants
and the respective decomposition of the energy-momentum tensor.

Assuming a one-component fluid with the four-velocity $\tensor{u}{^{a}}$, such
that $u_a u^a=-1$,
the gradient of $\tensor{u}{^{a}}$ can be decomposed kinematically
\cite{ellis1969rel-cosm}:
\begin{align} \label{eq:geschw_grad_zerleg}
	\tensor{u}{_{a;b}}
	=
	\tensor{\omega}{_{a b}} + \tensor{\sigma}{_{a b}}
	+ \frac{\Theta}{3}\, \tensor{h}{_{a b}}
	- \tensor{\dot{u}}{_a}\, \tensor{u}{_b} \, .
\end{align}
Thus, rotation, shear, acceleration and expansion as well as the scalar
quantities of rotation and acceleration read
\begin{gather} \notag
	%{}&
	\tensor{\omega}{_{a b}}
	=
		\tensor*{h}{*^c_a}\, \tensor*{h}{*^d_b}\, \tensor{u}{_{[c;d]}}	
	=
		\tensor{u}{_{[a;b]}} + \tensor{\dot{u}}{_{[a}}\, \tensor{u}{_{b]}}, \quad 
		\tensor{\sigma}{_{a b}}	= \tensor*{h}{*^c_a}\, \tensor*{h}{*^d_b}\, \tensor{u}{_{(c;d)}}
		- \frac{\Theta}{3}\, \tensor{h}{_{a b}},  \\ 
	%{}&
	\tensor{\dot{u}}{_a} = \tensor{u}{_{a;b}}\, \tensor{u}{^b}, \quad \Theta =
\tensor{u}{^a_{;a}}, \quad 
\omega^2 = (1/2)\, \tensor{\omega}{_{\alpha \beta}}\, \tensor{\omega}{^{\alpha
\beta}}, \quad
\dot{u}^2 = \tensor{\dot{u}}{_\alpha}\, .
\end{gather}
Choosing
$\tensor{u}{^{a}} = \tensor*{\delta}{*_{0}^{a}} $,
these kinematic quantities are rewritten
in the tetrad representation and for the space-times
\eqref{eq:metric_allg_obuk} as follows 
(the subscript $\Arrowvert$ denotes the covariant tetrad derivative)
\begin{align}
\label{eq:rotation_ted}
		\tensor{\omega}{_{\alpha \beta}}
	= {}&
		\tensor{u}{_{[\alpha \Arrowvert \beta]}}
		+ \tensor{\dot{u}}{_{[\alpha}}\, \tensor{u}{_{\beta]}}
	= \frac{1}{2\, a}\, \tensor{\hat{C}}{^\gamma_{\beta \alpha}}\, \zeta_{\gamma 0},
	 \\
\label{eq:beschl_def} %\notag
		\tensor{\dot{u}}{_{\alpha}}
		={}&
		  u_{\alpha \Arrowvert \rho}\, u^\rho
		=
		\frac{\dot{a}}{a}\, \tensor*{h}{*_{\alpha}^0},\\
\label{eq:expansion}
	\Theta
	={}&
	\tensor{u}{^\mu_{\Arrowvert \mu}}
	=
	\tensor{\Omega}{_\mu^\mu_0}
	=
	3\, \frac{\dot{a}}{a} ,
\end{align}
where $\tensor{u}{_{\alpha}} = \tensor{\zeta}{_{\alpha
0}}$.

According to \cite{ellis1969rel-cosm}, the energy-momentum tensor can be
decomposed with respect to the timelike velocity field $\tensor{u}{_a}$,
\begin{align} \label{eq:thermo-equil_ei-tensor_zerleg}
	{}& \tensor{T}{_{a b}}
	=
	\rho\, \tensor{u}{_a}\, \tensor{u}{_b}
	+ p\, \tensor{h}{_{a b}}
	+ 2\, \tensor{u}{_{(a}} \, \tensor{q}{_{b)}}
	+ \tensor{\pi}{_{a b}} \, .
\end{align}
Here the quantities can be identified with the
appropriate projections,
$\rho = \tensor{T}{_{a b}}\, \tensor{u}{^a}\,
	\tensor{u}{^b}$ for the energy density,
$p = \frac{1}{3}\, \tensor{T}{_{a b}}\, \tensor{h}{^{a b}}$ for the isotropic pressure,
$\tensor{q}{_a} = - \tensor{T}{_{c b}}\, \tensor{u}{^b}\,
\tensor*{h}{*_{a}^{c}}$ for the heat-flux and
$\tensor{\pi}{_{a b}}
	=
	\tensor{T}{_{c d}}\, \tensor*{h}{*_{a}^{c}}\, \tensor*{h}{*_{b}^{d}}
	- p\, \tensor{h}{_{a b}}$ for the anisotropic pressure.

\section{Matter equations} \label{sec:matter-eqs}

The conditions for the temperature vector
$\tensor{u}{_b}\, T^{-1}$, being a CKV, are
\begin{align} \label{eq:konf-faktor1}
	\Phi =  \frac{2}{3}\, \frac{\Theta}{T} \,
\end{align}
and
\begin{align} \label{eq:gleichgewicht_konf-killing_proj-expansion}
	\Theta
	=
	3\, T\, \left(\frac{1}{T}\right)_{,0} \, .
\end{align}
Both can be found in \cite{OlivierDavis} (the latter reproduces the $0$-component of (\ref{BEDINGUNG})).
The two results \eqref{eq:konf-faktor1} and
\eqref{eq:gleichgewicht_konf-killing_proj-expansion}
are obtained by inserting $\tensor{u}{_b}\, T^{-1}$
into \eqref{eq:diffgeo_conform_killing} and multiplying this equation
by $\tensor{u}{^a}\, \tensor{u}{^b}$ and $\tensor{g}{^{a b}}$, respectively.

Furthermore, integration of \eqref{eq:gleichgewicht_konf-killing_proj-expansion} leads to an expression
for the temperature scalar,
\begin{align} \label{eq:temp1}
	T ={}& \frac{1}{\tensor[_{\scriptscriptstyle{T}}]{c}{}\, a} \, ,
\end{align}
and the conformal factor,
\begin{align} \label{eq:konf-faktor2}
	\Phi = 2\, \tensor[_{\scriptscriptstyle{T}}]{c}{}\, \dot{a} \, ,
\end{align}
where $\tensor[_{\scriptscriptstyle{T}}]{c}{}$ is the constant of integration.

The existence of the CKV
has far-reaching consequences for the geometry of the space-time
and, factoring in Einsteins field equations, for the matter.
By drawing on the Ricci identity for the CKV and the Bianchi identity
subsequently, we deduce a set of four propagation equations \cite{borzchrob2006,borzchrobunknown}.
The first two describe the evolution of the energy density $\rho$
and the isotropic pressure $p$:
\begin{align}  \label{eq:propag_eqs_cond_energy_dens}
	{}&- \frac{1}{2}\, \square\, \Phi\, + \ddot{\Phi} - \Phi_{;m}\, \tensor{\dot{u}}{^m}
	+ \frac{1}{2\, T}\, \left( 3\, \dot{p} + \dot{\rho} \right)
	+ \frac{\Theta \left( 3\, p + \rho \right)}{3\, T}
	= 0
\end{align}
and
\begin{align} \label{eq:propag_eqs_cond_druck}
	%{}& \text{und} \qquad
	3\, \square\, \Phi - \frac{\left( 3\, \dot{p} - \dot{\rho} \right)}{T}
	- \frac{2\, \Theta \left( 3\, p - \rho \right)}{3\, T}
	= 0 \, .
\end{align}
The other two equations describe the change of the heat-flux $\tensor{q}{_a}$,
\begin{align}  \label{eq:propag_eqs_heatflow}
	\tensor*{h}{*^b_a} \, \dot{q}_b
   ={}&
   T \, \dot{\Phi}_{,b}\, \tensor*{h}{*^b_a}
   - T \, \Phi_{,m} \, \tensor{\omega}{^m_a}
   - \frac{1}{3}\, T \, \Phi_{,m} \, \Theta\, \tensor*{h}{*^m_a}
   - \frac{2}{3} \, \Theta \, q_a
   - q^k \, \omega_{ka} \, ,
\end{align}
and the anisotropic pressure $\tensor{\pi}{_{ab}}$,
\begin{align} \notag \label{eq:propag_eqs_anisotrop-druck}
	\tensor*{h}{*^m_a} \, \tensor*{h}{*^b_c} \, \dot{\pi}_{bm}
   = {}&
   - \frac{T}{2} \, \tensor{h}{_a_c}\, \square\, \Phi
   - T\, \tensor*{h}{*^m_a} \, \tensor*{h}{*^b_c} \, \Phi_{,m\,;b}
    + \tensor{h}{_a_c}\, \frac{\dot{p}-\dot{\rho}}{2}
   + 2 \tensor{\pi}{^k_{(a}}\, \tensor{\omega}{_{c)k}} \\
   &- \frac{2\,\Theta}{3}\, \pi_{ac}
   + \frac{\Theta\, (p-\rho)}{3}\ h_{ac} \, .
\end{align}

\subsection{Solutions of the propagation equations}
\label{sec:matter-eqs_non-perfect-fluid}
The reformulation of the dynamic equations \eqref{eq:propag_eqs_cond_energy_dens} -
\eqref{eq:propag_eqs_anisotrop-druck} in terms of the space-times \eqref{eq:metric_allg_obuk}
 and some tedious algebra brings us to 
 a set of ordinary differential
equations which can be solved analytically.
% (for details see \cite{schatz2012dipl}).

To this end, we decouple \eqref{eq:propag_eqs_cond_energy_dens} and \eqref{eq:propag_eqs_cond_druck} to
\begin{align} \label{eq:propag_eqs_energy_dens}
		\dot{\rho}
   =%= {}&
    - \frac{2}{3}\, \Theta\, \rho
    - T\, \square\, \Phi\,
    - T\, \ddot{\Phi}
    + T\, \tensor{\Phi}{_{,m}}\, \tensor{\dot{u}}{^m} \,
\end{align}
and
\begin{align} \label{eq:propag_eqs_druck}
	 	\dot{p}
   =%= {}&
    - \frac{2}{3}\, \Theta\, p
    + \frac{2}{3}\, T\, \square\, \Phi\,
    - \frac{1}{3}\, T\, \ddot{\Phi}
    + \frac{1}{3}\, T\, \tensor{\Phi}{_{,m}}\, \tensor{\dot{u}}{^m} \, .
\end{align}
By means of \eqref{eq:metric_alter_tetrad} equations \eqref{eq:propag_eqs_energy_dens} and \eqref{eq:propag_eqs_druck} can be rewritten as
\begin{align} \label{eq:energy_dens_int1}
	 	{}&	\left(a^2\, \rho \right)_{,0}
	 	+ 2\, \dddot{a}\, a \left( \tensor{\zeta}{^{0 0}} +1 \right)
	 	+ 2\, \ddot{a}\, \dot{a}
	 		\left(
	 			2\, \tensor{\zeta}{^{0 0}} - 1
	 		\right)
	 	+ 2\, \ddot{a}\, \tensor{\hat{C}}{^\gamma_{\kappa \gamma}}\, \tensor{\zeta}{^{\kappa 0}}
	 = 0
\end{align}
and
\begin{align} \label{eq:druck_int1}
	 {}& \left(a^2\, p \right)_{,0}
	 	+ \frac{2}{3}\, \dddot{a}\, a \left( 1 - 2\, \tensor{\zeta}{^{0 0}} \right)
	 	- \frac{2}{3}\, \ddot{a}\, \dot{a}
	 		\left(
	 			7\, \tensor{\zeta}{^{0 0}} + 1
	 		\right)
	 	- \frac{4}{3}\, \ddot{a}\, \tensor{\hat{C}}{^\gamma_{\kappa \gamma}}\, \tensor{\zeta}{^{\kappa 0}}
	 = 0 \, .
\end{align}
Integrating \eqref{eq:energy_dens_int1} and \eqref{eq:druck_int1}
yields
\begin{align} \label{eq:energy_dens_sol_final1}
	 	\rho
		={}&
		- 2\, \left( \frac{\dot{a}}{a} \right)_{,0}
		\left( \tensor{\zeta}{^{0 0}} +1 \right)
		- 3\, \left( \frac{\dot{a}}{a} \right)^2\, \tensor{\zeta}{^{0 0}}
		 - 2\, \frac{\dot{a}}{a^2}\,
			\tensor{\hat{C}}{^\gamma_{\kappa \gamma}}\, \tensor{\zeta}{^{\kappa 0}}
		+ \frac{1}{a^2}\, \tensor[_\rho]{c}{}  \, 
\end{align}
and
\begin{align}  \label{eq:druck_sol_final1}
	   p
	  ={}&
	      \frac{2}{3}\, \left( \frac{\dot{a}}{a} \right)_{,0}
		\left( 2\, \tensor{\zeta}{^{0 0}} - 1 \right)
	      + 3\, \left( \frac{\dot{a}}{a} \right)^2\, \tensor{\zeta}{^{0 0}}
	      + \frac{4}{3}\, \frac{\dot{a}}{a^2}
		\tensor{\hat{C}}{^\gamma_{\kappa \gamma}}\, \tensor{\zeta}{^{\kappa 0}}
	      + \frac{1}{a^2}\, \tensor[_p]{c}{} \, ,
\end{align}
where the objects $\tensor[_\rho]{c}{}$ and $\tensor[_p]{c}{}$ represent the
summed constants of integration of the additive antiderivatives, see Appendix \ref{sec:appendix-const-of-int}.
For concrete cosmological or
astrophysical models, e.g. stars, the boundaries
of the respective integrals are
specified.

With the identity
 \begin{align} \label{eq:anisotrop-druck_id}
	  \tensor*{h}{*_{\mu}^{\rho}}
	  \left(
		  \tensor{\zeta}{^{\mu \kappa}}\, \tensor{\omega}{_{\alpha \kappa}}
		  +
		  \tensor*{h}{*_{\alpha}^{\nu}}\, \tensor{u}{^{\gamma}}\, \tensor{\Omega}{_{\gamma}^{\mu}_{\nu}}
	  \right)
	  = 0 \, 
  \end{align}
and \eqref{eq:rotation_ted}, the tetrad formulations for the propagation
equations of the heat-flux \eqref{eq:propag_eqs_heatflow} and the anisotropic pressure \eqref{eq:propag_eqs_anisotrop-druck} together with the
kinematic quantities \eqref{eq:rotation_ted} - \eqref{eq:expansion} 
yield the following integrable partial differential equations:
\begin{align}
    \left(\tensor{q}{_{\alpha}}\, a^2\right)_{,0}
    +
    2\,
    \left(
	    \ddot{a}\, \dot{a} - \dddot{a}\, a
    \right)	\,
    \tensor*{h}{*_{\alpha}^0}
    +
    \ddot{a}\, \tensor{\hat{C}}{^{\hat{\gamma}}_{\alpha \sigma}}\,
    \tensor{\zeta}{_{\hat{\gamma} 0}}\, \tensor{\zeta}{^{\sigma 0}}
    = 0
\end{align}
and
\begin{align} \notag
	{}&
	3 \left(a^2\, \tensor{\pi}{_{\alpha \gamma}}\right)_{,0}
	+
	2\, \left(
	   \ddot{a}\, \dot{a} - \dddot{a}\, a
	\right)
	\left(
		\tensor{h}{_{\alpha \gamma}}
		\left(  \tensor{\zeta}{^{0 0}} + 1  \right)
		\right.\\ \notag
		& \left.\vphantom{}
		-
		3\, \tensor*{h}{*_{\alpha}^0}\, \tensor*{h}{*_{\gamma}^0}		
	\right)
	+
	2\, \ddot{a}\,
	\tensor{\hat{C}}{^\rho_{\mu \kappa}}\, \tensor{\zeta}{^{\kappa 0}}
	\left(
		\tensor*{\delta}{*_{\rho}^\mu}\, \tensor{h}{_{\alpha \gamma}}
		-
		3\, \tensor*{\delta}{*_{(\alpha}^\mu}\, \tensor{h}{_{\gamma) \rho }}		
	\right) \\
    ={}& 0  \,
\end{align}
for the heat-flux and the anisotropic pressure, respectively.

Integration and reorganization of terms bring the wanted solutions
\begin{align} \label{eq:heatflow_propag_sol_final1}
   \tensor{q}{_{\alpha}}
    =
    2\, \left( \frac{\dot{a}}{a} \right)_{,0} \tensor*{h}{*_{\alpha}^{0}}
    +
    \frac{\dot{a}}{a^2}\, \tensor{\hat{C}}{^{\gamma}_{\sigma \alpha}}\,\tensor{\zeta}{_{\gamma 0}}\,
    \tensor{\zeta}{^{\sigma 0}}
    +
    \frac{1}{a^2}\, \tensor[_q]{c}{_\alpha}
\end{align}
and
\begin{align} \notag \label{eq:anisotrop_druck_sol_final1}
	\tensor{\pi}{_{\alpha \beta}}
	={}&
	\frac{2}{3}\,
	\left( \frac{\dot{a}}{a} \right)_{,0} \tensor*{h}{*_{\alpha}^{0}}
	\left(
		\tensor{h}{_{\alpha \beta}} \left( \tensor{\zeta}{^{0 0}} +1 \right)
		-
		3\, \tensor*{h}{*_\alpha^0}\, \tensor*{h}{*_\beta^0}
	\right) \\
	&+
	\frac{2}{3}\, \frac{\dot{a}}{a^2}\,
	\tensor{\hat{C}}{^\rho_{\kappa \mu}}\, \tensor{\zeta}{^{\kappa 0}}
	\left(
		\tensor*{\delta}{*_{\rho}^\mu}\, \tensor{h}{_{\alpha \beta}}
		-
		3\, \tensor*{\delta}{*_{(\alpha}^\mu}\, \tensor{\zeta}{_{\beta) \rho }}				
	\right)
	+
	\frac{1}{a^2}\, \tensor[_\pi]{c}{_{\alpha \beta}} \, ,
\end{align}
where, similarly to the case of the energy density
\eqref{eq:energy_dens_sol_final1} and the isotropic pressure
\eqref{eq:druck_sol_final1} above, the objects $\tensor[_q]{c}{_\alpha}$
and $\tensor[_\pi]{c}{_{\alpha \gamma}}$ represent the
constants of integration
(see Appendix \ref{sec:appendix-const-of-int}).

With the help of the kinematic quantities
\eqref{eq:rotation_ted}-\eqref{eq:expansion} and
\eqref{eq:acc_div}-\eqref{eq:div_rot} of Appendix.
\ref{sec:apendix_kin-rels} the solutions
\eqref{eq:energy_dens_sol_final1},
 \eqref{eq:druck_sol_final1}, \eqref{eq:heatflow_propag_sol_final1}
 and \eqref{eq:anisotrop_druck_sol_final1} can be rewritten as follows:
\begin{align} \label{eq:energy_dens_sol_final2}
    \rho
   ={}&
    \frac{1}{3}\, \Theta^2
    + 3\, \dot{u}^2
    - 2\, \tensor{\dot{u}}{^{\gamma}_{\Arrowvert \gamma}}
    + T^2\, \tensor[_{\scriptscriptstyle{T}}]{c}{}^2\, \tensor[_\rho]{c}{} \,
    ,\\
%\end{align}
%
%\begin{align} 
\label{eq:druck_sol_final2}
    p
   ={}&
    - \frac{2}{3}\, \dot{\Theta}
    - \frac{1}{3}\, \Theta^2
    - \dot{u}^2
    + \frac{4}{3}\, \tensor{\dot{u}}{^{\gamma}_{\Arrowvert \gamma}}
    + T^2\, \tensor[_{\scriptscriptstyle{T}}]{c}{}^2\, \tensor[_p]{c}{}\, ,\\
%\end{align}
%
%\begin{align} 
\label{eq:heatflow_propag_sol_final2}
      \tensor{q}{_{\alpha}}
    ={}&
      \frac{2}{3}\, \dot{\Theta}\, \tensor*{h}{*_{\alpha}^{0}}
      +
      2\, \tensor{\omega}{_{\alpha \gamma}}\, \tensor{\dot{u}}{^{\gamma}}
      +
      T^2\, \tensor[_{\scriptscriptstyle{T}}]{c}{}^2\,
      \tensor[_{q}]{c}{_{\alpha}}\,
\end{align}
and
\begin{align} \notag \label{eq:anisotrop_druck_sol_final2}
      \tensor{\pi}{_{\alpha \beta}}
    ={}&
     	- \frac{2}{3}\, \dot{\Theta}\, \tensor*{h}{*_{\alpha}^{0}}\, \tensor*{h}{*_{\beta}^{0}}
      -
      2\, \dot{u}^2\, \tensor{h}{_{\alpha \beta}}
      +
      \frac{2}{3}\, \tensor{\dot{u}}{^{\gamma}_{\Arrowvert \gamma}}\,
      \tensor{h}{_{\alpha \beta}}\\
      &-
      2\, T\, \tensor{\dot{u}}{^{\kappa}}\, \tensor{\hat{C}}{^{\rho}_{\kappa
      (\alpha}}\, \tensor{h}{_{\beta) \rho}}
      +
      T^2\, \tensor[_{\scriptscriptstyle{T}}]{c}{}^2\,
      \tensor[_{\pi}]{c}{_{\alpha \beta}}\,
\end{align}
or, in terms of purely kinematic quantities,
\begin{align} \notag \label{eq:anisotrop_druck_sol_final2_test}
      \tensor{\pi}{_{\alpha \beta}}
    ={}&
     \frac{2}{3}\, \dot{\Theta}
      \left(
	2\, \tensor*{h}{*_{(\alpha}^{0}}\, \tensor{u}{_{\beta)}}
      	-3\, \tensor*{h}{*_{\alpha}^{0}}\, \tensor*{h}{*_{\beta}^{0}}
      \right)
      +
      \frac{2}{3}
      \left(
      	\tensor{\dot{u}}{^{\gamma}_{\Arrowvert \gamma}}
      	- 9\, \dot{u}^2
      \right)
      \tensor{h}{_{\alpha \beta}} \\\notag
      &+
      4\, \tensor{\dot{u}}{_{(\alpha \Arrowvert \beta)}}
      -
       \frac{4}{3}\, \Theta\,
      	\tensor{\dot{u}}{_{(\alpha}}\, \tensor{u}{_{\beta)}}
      +
      4\, \tensor{\omega}{_{\kappa (\alpha}}\, \tensor{u}{_{\beta)}}\,
      \tensor{\dot{u}}{^{\kappa}} \\
      &+
      4\, \tensor{\dot{u}}{_{\alpha}}\, \tensor{\dot{u}}{_{\beta}}
      -
      4\, \dot{u}^2\,
      \tensor{u}{_{\alpha}}\, \tensor{u}{_{\beta}}
      +
      T^2\, \tensor[_{\scriptscriptstyle{T}}]{c}{}^2\,
      \tensor[_{\pi}]{c}{_{\alpha \beta}}\, .
\end{align}
The expressions \eqref{eq:energy_dens_sol_final2} and
\eqref{eq:druck_sol_final2} can be understood as generalized
Friedmann equations.

According to
\eqref{eq:thermo-equil_ei-tensor_zerleg}, we can now reconstruct the
energy-momentum tensor by inserting the four solutions above:
\begin{align} \notag \label{eq:re_ei_tensor_final1} \notag
    \tensor{T}{_{\alpha \beta}}
  ={}&
    - \frac{2}{3}\, \dot{\Theta}
    \left(
      \tensor{\zeta}{_{\alpha \beta}}
      +
      \tensor*{\delta}{*_\alpha ^0}\, \tensor*{\delta}{*_\beta ^0}
    \right)
    +
    \frac{1}{3}\,
    \left(
      6\, \tensor{\dot{u}}{^{\gamma}_{\Arrowvert \gamma}}
      - \Theta^2
      - 9\, \dot{u}^2
    \right)
    \tensor{\zeta}{_{\alpha \beta}}\\
    &-
    2\, T\, \tensor[_{\scriptscriptstyle{T}}]{c}{}\,
    \tensor{\dot{u}}{^{\kappa}}\, \tensor{\hat{C}}{^{\rho}_{\kappa (\alpha}}\, \tensor{\zeta}{_{\beta) \rho}}
    +
    T^2\, \tensor[_{\scriptscriptstyle{T}}]{c}{}^2\,
    \tensor[_{\scriptscriptstyle{EI}}]{c}{_{\alpha \beta}} \, ,
\end{align}
or
\begin{align} \notag \label{eq:re_ei_tensor_final2}
	\tensor{T}{_{\alpha \beta}}
  ={}&
	\frac{2}{3}\, \dot{\Theta}\,
	\left(
		2\, \tensor*{h}{*_{(\alpha}^{0}}\, \tensor{u}{_{\beta)}}
		-
		3\, \tensor*{h}{*_{(\alpha}^{0}}\, \tensor*{h}{*_{\beta)}^{0}}
		-
		\tensor{h}{_{\alpha \beta}}
	\right)
	-
	\dot{u}^2
	\left(
		7\, \tensor{h}{_{\alpha \beta}}
		+
		\tensor{u}{_{\alpha}}\, \tensor{u}{_{\alpha}}
	\right) \\
	&+
	\frac{1}{3}
	\left(
		\tensor{\dot{u}}{^{\gamma}_{\Arrowvert \gamma}}
		-
		\Theta^2
	\right)
	\tensor{\zeta}{_{\alpha \beta}}
	+
	4
	\left(
		\tensor{\dot{u}}{_{(\alpha \Arrowvert \beta)}}
		+
		\tensor{\dot{u}}{_{\alpha}}\, \tensor{\dot{u}}{_{\beta}}
	\right)
	+
	T^2\, \tensor[_{\scriptscriptstyle{T}}]{c}{}^2\,
	\tensor[_{\scriptscriptstyle{EI}}]{c}{_{\alpha \beta}}\, ,
\end{align}
where
\begin{align} \label{eq:int_const_general_ei}
	\tensor[_{\scriptscriptstyle{EI}}]{c}{_{\alpha \beta}}
	=
	\tensor[_{\rho}]{c}{}\, \tensor{\zeta}{_{\alpha 0}}\, \tensor{\zeta}{_{\beta 0}}
	+
	\tensor[_{p}]{c}{}\, \tensor{h}{_{\alpha \beta}}
	+
	2\, \tensor[_q]{c}{_{(\alpha}}\, \tensor{\zeta}{_{\beta) 0}}
	+
	\tensor[_\pi]{c}{_{\alpha \beta}}\, .
\end{align}
In order to verify the consistency of the solutions
\eqref{eq:energy_dens_sol_final2} - \eqref{eq:anisotrop_druck_sol_final2} with Einstein's
field equations and in order to recover special cases, the
constants of integration are
determined by the calculations of Appendix \ref{sec:appendix-const-of-int}. The matter equations
\eqref{eq:energy_dens_sol_final2} - \eqref{eq:anisotrop_druck_sol_final2} then
take the exclusively kinematic forms
\begin{align} \label{eq:energy_dens_sol_final-re}
    \rho
  ={}&
    \frac{1}{3}\, \Theta^2
    + 3\, \dot{u}^2
    - 2\, \tensor{\dot{u}}{^{\gamma}_{\Arrowvert \gamma}}
    -  \frac{1}{2}\, \tilde{R}\, T^2\,
    \tensor[_{\scriptscriptstyle{T}}]{c}{}^2\,  + 2\, \omega^2 \, ,\\
%\end{align}
%
%\begin{align} 
\label{eq:druck_sol_final-re}
    p
   ={}&
    - \frac{2}{3}\, \dot{\Theta}
    - \frac{1}{3}\, \Theta^2
    - \dot{u}^2
    + \frac{4}{3}\, \tensor{\dot{u}}{^{\gamma}_{\Arrowvert \gamma}}
    + \frac{1}{6}\, \tilde{R}\, T^2\,
    \tensor[_{\scriptscriptstyle{T}}]{c}{}^2\, + \frac{2}{3}\, \omega^2\, ,\\
%\end{align}
%
%\begin{align} 
\label{eq:heat_flow_sol_final-re}
      \tensor{q}{_{\alpha}}
    ={}&
      \frac{2}{3}\, \dot{\Theta}\, \tensor*{h}{*_{\alpha}^{0}}
      +
      \tensor{\omega}{_{\alpha \gamma}}\, \tensor{\dot{u}}{^{\gamma}}
      +
      \tensor{\omega}{^{\tau \gamma}_{\Arrowvert \gamma}}\,
      \tensor{h}{_{\tau \alpha}}\,,\\
%\end{align}
%
%and
%
%\begin{align} 
\notag \label{eq:anisotrop_druck_sol_final2-ext}
      \tensor{\pi}{_{\alpha \beta}}
    ={}&
      \frac{2}{3}\, \dot{\Theta}
      \left(
      	2\, \tensor*{h}{*_{(\alpha}^{0}}\, \tensor{u}{_{\beta)}}
      	-
      	3\, \tensor*{h}{*_{\alpha}^{0}}\, \tensor*{h}{*_{\beta}^{0}}
      \right)
      -
      2
      \left(
      	 2\, \dot{u}^2
      	+ \omega^2
      \right)
      \tensor{u}{_{\alpha}}\, \tensor{u}{_{\beta}}\\\notag
      &+
      \frac{1}{3}
      \left(
      	2\, \tensor{\dot{u}}{^{\gamma}_{\Arrowvert \gamma}}
      	-
      	18\, \dot{u}^2
      	-
      	2\, \omega^2
      	+
      	\tilde{R}\, T^2\, \tensor[_{\scriptscriptstyle{T}}]{c}{}^2
      \right)
      \tensor{h}{_{\alpha \beta}} \\\notag
      &-
      \tensor{\tilde{R}}{_{\alpha \beta}}\, T^2\,
      \tensor[_{\scriptscriptstyle{T}}]{c}{}^2\,
      -
      \frac{4}{3}\, \Theta\,
      	\tensor{\dot{u}}{_{(\alpha}}\, \tensor{u}{_{\beta)}}
      +
      4\, \tensor{\dot{u}}{_{(\alpha \Arrowvert \beta)}}
      +
      4\, \tensor{\dot{u}}{_{\alpha}}\, \tensor{\dot{u}}{_{\beta}}\\
      &+
      2\, \tensor{\omega}{_{\kappa (\alpha}}\, \tensor{u}{_{\beta)}}\,
      \tensor{\dot{u}}{^{\kappa}}
      -
      2\, \tensor{\omega}{^{\tau \mu}_{\Arrowvert \mu}}\, \tensor{h}{_{\tau (\alpha}}\,
      \tensor{u}{_{\beta)}}\, ,
\end{align}
in which for the latter, $\tensor{\pi}{_{\alpha \beta}}$, relation
\eqref{eq:grad_acc} was used in addition. By multiplying these expressions with the tetrads
one obtains the coordinate representation without any additional terms.

Notice, that \eqref{eq:energy_dens_sol_final-re} and
\eqref{eq:druck_sol_final-re} satisfy the Raychaudhuri equation:
\begin{align}
	\rho + 3\, p
	  =
	     - 2\, \dot{\Theta}
	    -
	    \frac{2}{3}\, \Theta^2\,
	   +
	    4\, \omega^2
	   + 2\, \tensor{\dot{u}}{^{a}_{; a}}\, .
\end{align}
  \subsection{Special cases} \label{sec:matter-eqs_special-cases}
\subsubsection{Non-tilted models}\label{sec:matter-eqs_non-tilted_models}
The non-tilted limit ($\tensor*{h}{*_{\alpha}^{0}} = 0$) leads to purely
expanding models, i.e., those with vanishing rotation and acceleration.
In this case also the coefficients $\tensor{n}{_{\hat{a}}}$ of
\eqref{eq:metric_allg_obuk} become zero,
so that the space-times are conformally static \cite{coley1994, perlick2004}.

Following up on this premise,
  \eqref{eq:energy_dens_sol_final-re} and \eqref{eq:druck_sol_final-re}
  turn into the equations
\begin{align} \label{eq:rel_sol_perf-fluid_energy_friedmann2}
		\rho
	 		={}&
	 			3\, \left(\frac{\dot{a}}{a} \right)^2
		 		- \frac{\tilde{R}}{2\, a^2}
\end{align}	
and
\begin{align} \label{eq:rel_sol_perf-fluid_druck_friedmann2} %\notag
	 	p
	  =
	    - \left( \frac{\dot{a}}{a} \right)^2
		- 2\, \frac{\ddot{a}}{a}
	     + \frac{\tilde{R}}{6\, a^2} \, .
\end{align}
These correspond to the Friedmann equations
\begin{align} \label{eq:rel_sol_perf-fluid_goenner_energy_friedmann}
	\frac{1}{3}\, \rho
	=
	\left( \frac{\dot{a}}{a} \right)^2
	+ \frac{k}{a^2}
\end{align}
and
\begin{align} \label{eq:rel_sol_perf-fluid_goenner_druck_friedmann}
	- p
	=
	2\, \frac{\ddot{a}}{a}
	+ \left( \frac{\dot{a}}{a} \right)^2
	+ \frac{k}{a^2} \, , 
\end{align}
if the curvature parameter $k$ and the Ricci scalar $\tilde{R}$ of the
3-dimensional Bianchi spaces  are related by
\begin{align}
	k = - \frac{1}{6}\, \tilde{R} \, .
\end{align}
This result is in accordance with \cite[p.474]{rebhan2012}.
The two constants of integration are then related by
\begin{align}
	\tensor[_p]{c}{} = - \frac{1}{3}\, \tensor[_\rho]{c}{} = \frac{1}{6}\,
	\tilde{R} \, .
\end{align}

Furthermore, the heat-flux \eqref{eq:heat_flow_sol_final-re} is identically zero,
while for the anisotropic pressure \eqref{eq:anisotrop_druck_sol_final2-ext} one
gets
\begin{align} \label{eq:non-tilted_anisotrop-press}
	\tensor{\pi}{_{\alpha \beta}}
	= \tensor[_\pi]{c}{_{\alpha \beta}}\, T^2\,
	\tensor[_{\scriptscriptstyle{T}}]{c}{}^2\,
	=
	- \tensor{\tilde{R}}{_{\alpha \beta}}\, T^2\,
	\tensor[_{\scriptscriptstyle{T}}]{c}{}^2\,  + \frac{1}{3}\, \tilde{R}\,
	T^2\, \tensor[_{\scriptscriptstyle{T}}]{c}{}^2\, \tensor{h}{_{\alpha \beta}}\,
	.
\end{align}
According to Sec. 4 of \cite{coley1994}, the non-perfect fluid models
investigated here can be subdivided into three further classes.
In detail, this amounts to determining the number of distinct eigenvalues of
the anisotropic pressure \eqref{eq:non-tilted_anisotrop-press}:
If $\tensor{\pi}{_{\alpha \beta}}$ has three different eigenvalues,
the space-time is of Petrov-type I. In the case of two different eigenvalues, one
obtains Petrov-type D. Finally, if there is only one eigenvalue, 
it can only be
zero and results in $\tensor{\pi}{_{\alpha \beta}}$ vanishing identically.
Therefore, only this latter case is in general a sufficient condition for obtaining
perfect fluid Friedmann models.

An example for a non-tilted space-time subclass of \eqref{eq:metric_allg_obuk}
which does not contain Friedmann models is provided by the Bianchi-type IV metric
\begin{align} \notag \label{eq:bianchi4-general-metric}
	ds^2={}&
	-dt^2
	+
	a^2
	\left(
		dx^2\,\tensor{\beta}{_{1 1}}
		+
		2\,dx\,e^x
		\left(
			dz\,\tensor{\beta}{_{1 3}}
			+
			dy
			\left(
				\tensor{\beta}{_{1 2}}	
				+
				x\,\tensor{\beta}{_{1 3}}
			\right)
		\right)
		\right.\\ \notag
		& \left.\vphantom{}
		+
		e^{2\,x}
		\left(
			dz^2\,\tensor{\beta}{_{3 3}}
			+
			2\,dy\,dz
			\left(
				\tensor{\beta}{_{2 3}}
				+
				x\,\tensor{\beta}{_{3 3}}
			\right)
			\right.
			\right.\\
			& \left.\vphantom{}
			\left.\vphantom{}		
			+
			dy^2
			\left(
				\tensor{\beta}{_{2 2}}
				+
				2\,x\,\tensor{\beta}{_{2 3}}
				+
				x^2\,\tensor{\beta}{_{3 3}}
			\right)
		\right)
	\right)
\end{align}
with canonic coordinates $(x^0 \rightarrow t,\, x^1 \rightarrow x,\, x^2
\rightarrow y,\, x^3 \rightarrow z)$ and undetermined constants
$\beta_{\hat{a} \hat{b}}$ introduced in
\eqref{eq:metric_allg_obuk}.
Since here one finds three distinct
eigenvalues for \eqref{eq:non-tilted_anisotrop-press}, the models
\eqref{eq:bianchi4-general-metric} are of Petrov-type I.

For the particular case $\beta_{\hat{a} \hat{b}} = \mathbb{1}$ the metric \eqref{eq:bianchi4-general-metric}
takes the form
\begin{align}  \label{eq:bianchi4-spec-metric}
	ds^2={}&
	-dt^2
	+
	a^2
	\left(
		dx^2
		+
		e^{2\,x}
		\left(
			dz^2\,
			+
			2\, x\, dy\, dz
			+
			dy^2
			\left(
				1
				+
				x^2
			\right)
		\right)
	\right)
\end{align}
which coincides exactly with the example discussed in \cite{coley1994} (see
(4.9) therein).

\subsubsection{Stationary models}\label{sec:matter-eqs_stationary-models}
We check the consistency of our results with the stationary
limit as to \cite{obukhov2000,
obukhov1990bianchi_list}.
Accordingly, for vanishing expansion and generally non-trivial
rotation, the equations
\eqref{eq:energy_dens_sol_final-re} and \eqref{eq:druck_sol_final-re}
reduce to
\begin{align}  \label{eq:energy-dens_godel}
     \rho =  \tensor[_\rho]{c}{}\, \tensor[_{\scriptscriptstyle{T}}]{c}{}^2\,
     T^2 =
      - \frac{1}{2}\, \tilde{R}\, T^2\,
      \tensor[_{\scriptscriptstyle{T}}]{c}{}^2\, + 2\, \omega^2 \,
\end{align}
and
\begin{align} \label{eq:druck_godel}
     p = \tensor[_p]{c}{}\, \tensor[_{\scriptscriptstyle{T}}]{c}{}^2\, T^2
   =
      \frac{1}{6}\, \tilde{R}\, T^2\, \tensor[_{\scriptscriptstyle{T}}]{c}{}^2\,
      + \frac{2}{3}\, \omega^2\, ,
\end{align}
so that
\begin{align} \label{eq:pot_eq_state_godel}
	p = \frac{\tensor[_p]{c}{}}{\tensor[_\rho]{c}{}}\, \rho
	 = const \, .
\end{align}
In the perfect fluid limit with vanishing heat-flux and anisotropic
pressure, one obtains from \eqref{eq:heat_flow_sol_final-re} the condition
\begin{align} \notag \label{eq:int_const_heat-flow_godel}
	\tensor{q}{_{\alpha}} = \tensor[_q]{c}{_\alpha}\,
	\tensor[_{\scriptscriptstyle{T}}]{c}{}^2\, T^2
	={}&
	    \tensor{\omega}{^{\kappa \mu}_{\Arrowvert \mu}}\,
		\tensor{h}{_{\alpha \kappa}}\\
	={}& 0	\, 	
\end{align}
and from \eqref{eq:anisotrop_druck_sol_final2-ext}
\begin{align} \notag \label{eq:int_const_anisotrop-druck_godel}
      \tensor{\pi}{_{\alpha \beta}} ={}& \tensor[_\pi]{c}{_{\alpha \beta}}\,
      \tensor[_{\scriptscriptstyle{T}}]{c}{}^2\, T^2\\\notag
      ={}&
	- \tensor{\tilde{R}}{_{\alpha \beta}}\, T^2\,
	\tensor[_{\scriptscriptstyle{T}}]{c}{}^2\,
	+
	\frac{1}{3}\, \tilde{R}\, T^2\, \tensor[_{\scriptscriptstyle{T}}]{c}{}^2\,
	\tensor{h}{_{\alpha \beta}} -
	\frac{2}{3}\, \omega^2
	\left(
	  3\, \tensor{u}{_{\alpha}}\, \tensor{u}{_{\beta}}
	 +
	  \tensor{h}{_{\alpha \beta}}
	\right)\\\notag
	&-
	2
	  \tensor{\omega}{^{\tau \mu}_{\Arrowvert \mu}}\, \tensor{h}{_{\tau (\alpha}}\, \tensor{u}{_{\beta)}}\\
    ={}& 0 \, .
\end{align}
As a more concrete \textit{ansatz} we choose a Bianchi-type III subclass of the
space-times \eqref{eq:metric_allg_obuk},
\begin{align} \notag \label{eq:godel-type_sptime}
  (ds)^2 ={}& (dt)^2 - 2\, \sqrt{\Sigma}\, a\, e^{M\, x^1}\, dt\, dx^2
		- a^2\, \left(
			(dx^1)^2
			\vphantom{\frac{}{}}\right.\\
			& \left.\vphantom{\frac{}{}}
			+ K\, e^{2\, M\, x^1}\, (dx^2)^2 + (dx^3)^2
		\right)\, ,
\end{align}
where $\tensor{\nu}{_{\hat{a}}} = \left( 0, \sqrt{\Sigma},0 \right)$,
$\tensor{\beta}{_{\hat{a} \hat{b}}} = \textrm{diag}\left( 1,K,1 \right)$ and
$\tensor{e}{^{\hat{\mu}}_{\hat{a}}} = \textrm{diag}\left( 1,e^{M\, \tensor{x}{^1}},1 \right)$ with $K$, $M$ and $\Sigma$ being constant.
Admitting in general non-vanishing rotation and expansion,
this metric is also denoted as the G\"odel-type model (see \cite{obukhov1990bianchi_list,obukhov2000}).
By this choice, the heat-flux \eqref{eq:int_const_heat-flow_godel} vanishes
identically, whilst the anisotropic pressure condition \eqref{eq:int_const_anisotrop-druck_godel} holds only for at least
either of the two relations,
\begin{align} \label{eq:cond_class_godel_consts}
  K = - \frac{\Sigma}{2} \qquad \text{or} \qquad M = 0\, .
\end{align}
Furthermore, one has
\begin{align}
  \tilde{R} = \frac{M^2 \left( 4\, K + 3\, \Sigma \right)}{2\, \left( K + \Sigma \right) }
  \quad
  \text{and}
  \quad
  \omega^2 = \frac{M^2 \Sigma}{4\, a^2\, \left( K + \Sigma \right) }
\end{align}
or, by \eqref{eq:cond_class_godel_consts} respectively,
\begin{align}
  \tilde{R} = M^2
  \quad
  \text{and}
  \quad
  \omega^2 = \frac{M^2}{2\, a^2\, }\, .
\end{align}
This yields $\tensor[_p]{c}{} = \tensor[_\rho]{c}{}$ and thus for
expression \eqref{eq:pot_eq_state_godel},
\begin{align}
	p = \rho = \frac{1}{2}\, \tilde{R}\, T^2\,
	\tensor[_{\scriptscriptstyle{T}}]{c}{}^2
	= \omega^2 .
\end{align}
According to, e. g., \cite{stephani1988} this is just
the equation of state of the classical G\"odel space-time. Indeed, in
\cite{obukhov1990bianchi_list} it is stated, that $K = - (1/2)\, \Sigma$
yields closed timelike curves.
%
%
%\newpage
\section{Discussion}\label{sec:Interpretation}

In this paper, we considered homogeneous and conform-stationary space-times
\eqref{eq:metric_allg_obuk} with Bianchi group invariance and an arbitrary
matter source, which allows for generally tilted models.

By solving the propagation equations
\eqref{eq:propag_eqs_cond_energy_dens} - \eqref{eq:propag_eqs_anisotrop-druck},
we deduce explicit expressions for the energy density
\eqref{eq:energy_dens_sol_final1}, the isotropic pressure
\eqref{eq:druck_sol_final1}, the heat-flux
\eqref{eq:heatflow_propag_sol_final1} and the anisotropic pressure
\eqref{eq:anisotrop_druck_sol_final1} in terms of the scale factor, the tetrad
components \eqref{eq:metric_alter_tetrad} and the structure constants. These results are rewritten
in terms of the kinematic quantities, as to be found in
\eqref{eq:energy_dens_sol_final2}, \eqref{eq:druck_sol_final2},
\eqref{eq:heatflow_propag_sol_final2} and
\eqref{eq:anisotrop_druck_sol_final2_test}, and are combined to
the energy-momentum tensor, \eqref{eq:re_ei_tensor_final1} or
\eqref{eq:re_ei_tensor_final2}. Similar equations are \textit{ad hoc} 
assumed
in \cite{NOJIRI1, NOJIRI2, CAPOZIELLO2006, SEBASTIANI2014} 
in order to solve problems arising during cosmological evolution for different reasons.

In addition to the Raychaudhuri equation and the
other propagation and constraint equations
(see, e.g., \cite{ellis1969rel-cosm, chrobok2004}),
 we obtain equations in which the expressions for the matter content are
 decoupled and independent of higher derivatives of the kinematic
 quantities (except for the expansion and acceleration) or depending
 on the electric part of the Weyl tensor. Especially, no equations
 of state or further thermodynamic relations have to be assumed to
 arrive at these results. Here it should be emphasized that the
vanishing shear does not necessarily imply a zero anisotropic pressure as
required by linear thermodynamics. It should also be pointed out that more-component
 fluids or a cosmological constant can easily be included.

The equations \eqref{eq:energy_dens_sol_final1}, 
\eqref{eq:druck_sol_final1}, 
\eqref{eq:heatflow_propag_sol_final1} and 
\eqref{eq:anisotrop_druck_sol_final1}
represent a class of models which does not only contain physically relevant
space-times.
To take into account well-motivated (energy) conditions or global
aspects (as considered in \cite{HARRIS}) which should provide broader
restrictions, 
is therefore a subject of future research.

Moreover, inspection of the equations
 \eqref{eq:energy_dens_sol_final2}, \eqref{eq:druck_sol_final2},
 \eqref{eq:heatflow_propag_sol_final2} and
 \eqref{eq:anisotrop_druck_sol_final2_test}  underlines that further
 thermodynamic assumptions like an equation of state, Fourier's law,
 Cauchy's law or expressions from extended thermodynamics,
 will further restrict possible solutions.
This becomes manifest, if one rewrites \eqref{eq:energy_dens_sol_final1},
 	\eqref{eq:druck_sol_final1}, \eqref{eq:heatflow_propag_sol_final1}
 	and \eqref{eq:anisotrop_druck_sol_final1}
 with the help of equation
 	\eqref{eq:gleichgewicht_konf-killing_proj-expansion}:
\begin{align} \notag
%energ. dens.
	\rho
	={}&
	-
	\left(
		\frac{\dot{T}}{T}
	\right)^2
	\left(
		5\, \tensor{\zeta}{^{0 0}} + 2
	\right)
	+
	2\, \frac{\ddot{T}}{T}
	\left(
		\tensor{\zeta}{^{0 0}} + 1
	\right)
	+
	2\, \dot{T}\, \tensor[_{\scriptscriptstyle{T}}]{c}{}\, \tensor{\hat{C}}{^{\gamma}_{\kappa
	\gamma}}\,\tensor{\zeta}{^{\kappa 0}} \\
	\label{eq:HEINZ1}
	&+
	T^2\, \tensor[_{\scriptscriptstyle{T}}]{c}{}^2\, \tensor[_{\rho}]{c}{}\, ,
	\\\notag
% isotop. press
	p
	={}&
	\frac{1}{3}
	\left(
		\frac{\dot{T}}{T}
	\right)^2
	\left(
		13\, \tensor{\zeta}{^{0 0}} - 2
	\right)
	-
	\frac{2}{3}\, \frac{\ddot{T}}{T}
	\left(
		2\, \tensor{\zeta}{^{0 0}} - 1
	\right)
	-
	\frac{4}{3}\, \dot{T}\, \tensor[_{\scriptscriptstyle{T}}]{c}{}\,
	\tensor{\hat{C}}{^{\gamma}_{\kappa \gamma}}\,\tensor{\zeta}{^{\kappa 0}} \\
	\label{eq:PETER1}
	&+
	T^2\, \tensor[_{\scriptscriptstyle{T}}]{c}{}^2\, \tensor[_{p}]{c}{}\, ,\\
% heat flow
	 \label{eq:WILLI1}
	\tensor{q}{_{\alpha}}
	={}&
	2
	\left(
		\frac{\dot{T}^2 - \ddot{T}\, T}{T^2}
	\right)
	\tensor*{h}{*^{0}_{\alpha}}
	-
	\dot{T}\, \tensor[_{\scriptscriptstyle{T}}]{c}{}\,
	\tensor{\hat{C}}{^{\gamma}_{\sigma \alpha}}\, \tensor{\zeta}{_{\gamma
	0}}\, \tensor{\zeta}{^{\sigma 0}}
	+
	T^2\, \tensor[_{\scriptscriptstyle{T}}]{c}{}^2\,
	\tensor[_{q}]{c}{_{\alpha}}\, ,\\
%\end{align}
%\begin{align} 	
% anisotrop. press.%
	\notag
	\tensor{\pi}{_{\alpha \beta}}
	={}&
	\frac{2}{3}
	\left(
		\frac{\dot{T}^2 - \ddot{T}\, T}{T^2}
	\right)
	\left(
		\tensor{h}{_{\alpha \beta}}
		\left(
			\tensor{\zeta}{^{0 0}} + 1
		\right)
		-
		3\, \tensor*{h}{*^{0}_{\alpha}}\, \tensor*{h}{*^{0}_{\beta}}
	\right) \label{eq:GUSTAV1} \\
	&-
	\frac{2}{3}\,
	\dot{T}\, \tensor[_{\scriptscriptstyle{T}}]{c}{}\,
	\tensor{\hat{C}}{^{\rho}_{\kappa \mu}}\, \tensor{\zeta}{^{\kappa 0}}
	\left(
		\tensor*{\delta}{*^{\mu}_{\rho}}\, \tensor{h}{_{\alpha \beta}}
		-
		3\, \tensor*{\delta}{*^{\mu}_{(\alpha}}\, \tensor{\zeta}{_{\beta) \rho}}
	\right)
	+
	T^2\, \tensor[_{\scriptscriptstyle{T}}]{c}{}^2\, \tensor[_{\pi}]{c}{_{\alpha
	\beta}} \, .
\end{align}
These equations describe the temperature dependence of the matter content which has to be fulfilled for the considered class of models.

The Expressions \eqref{eq:HEINZ1} and \eqref{eq:PETER1}
can be used to construct equations of state.
For instance, one can combine the two in such a way that the outcome does not
contain the structure constants:
\begin{align} \label{eq:HEINZ2}
p + 2\, \rho
	=
	3
	\left(
		\frac{\dot{T}}{T}
	\right)^2
	\left(
		\tensor{\zeta}{^{0 0}} - 2
	\right)
	+
	6\, \frac{\ddot{T}}{T}
	+
	T^2\, \tensor[_{\scriptscriptstyle{T}}]{c}{}^2
	\left(
		 2\, \tensor[_{\rho}]{c}{}
		 +
		 3\, \tensor[_{p}]{c}{}
	\right)
\end{align}
which is a possible equation of state for the considered
space-time class. This relation clearly shows that
the pressure has a difficult dependence on the temperature
and its first and second derivatives. Of course, $\rho$
has an explicit temperature dependence as given
in \eqref{eq:HEINZ1}, but assuming the validity of
simple equations of state, like $p(\rho)\propto \rho^{\alpha}$,
an effective fine-tuning has to be done in order to prevent an additional
temperature dependence of $p$.

It becomes obvious from \eqref{eq:WILLI1} and \eqref{eq:GUSTAV1} 
that the assumption of Fourier's or Cauchy's laws consequently generates additional strong restrictions 
on the space-time and its matter content.
The same is true for other \textit{ad hoc} introduced constitutive equations.
This includes non-linear ones like the heat-flux law of Israel-Stewart-type
\cite{israel1989fuid-intro} which is physically motivated by that it overcomes 
stability and causality problems arising in the linear case.
Our point, however, is to ask for those constitutive equations 
and equations of state respectively, which follow from the conservation laws in
a prescribed geometry and a given temperature field.
%,and these are shown to be the equations (81) and (82). 
Thus, it is not in the sense of the present consideration, to additionally
impose \textit{ad hoc } constitutive equations on \eqref{eq:WILLI1} and
\eqref{eq:GUSTAV1}.
In non-relativistic continuum thermodynamics the situation is different.
There, one has to complete the system of basic equations resting on the
conservation or balance equations by adding such \textit{ad hoc} relations
manually.
If \eqref{eq:WILLI1} and \eqref{eq:GUSTAV1} differ from those \textit{ansatzes}
made by hand, this can have a variety of reasons and implications. To call only one: 
If there were severe thermodynamic arguments for one of the linear 
or non-linear \textit{ad hoc} 
\textit{ansatzes}, e.g., for the heat-flux, one was obliged to ask
under which condition it is compatible with \eqref{eq:WILLI1}.

This view is reinforced by the results obtained in \cite{HERRERA}. There, it was
shown that in conform-stationary models the heat-flux must vanish for zero
anisotropic pressure and under the assumption of a heat-flux law of the
Israel-Stewart type.
An example for physical processes in which this does not hold (Landau damping)
is also provided in \cite{HERRERA}.

Moreover, the form of the expressions \eqref{eq:WILLI1} and \eqref{eq:GUSTAV1},
which denotes the modified laws of Fourier and Cauchy, is
pointing in a direction that is to be found in various formulations of extended thermodynamics \cite{israel1989fuid-intro, ruggeri1993}. This becomes 
evident if one rewrites \eqref{eq:GUSTAV1} with the help
of \eqref{eq:HEINZ1} and \eqref{eq:WILLI1}:
\begin{align} \notag
%anisotrop. press.
   \tensor{\pi}{_{\alpha \beta}}
   	={}&
   	\frac{T^2}{2\, \left( \ddot{T}\, T - \dot{T}^2 \right) }
   	\left(
   		\tensor{q}{_{\alpha}}
   		+
   		\dot{T}\, \tensor[_{\scriptscriptstyle{T}}]{c}{}\,
   		\tensor{\hat{C}}{^{\gamma}_{\sigma \alpha}}\, \tensor{\zeta}{_{\gamma 0}}\,
   		\tensor{\zeta}{^{\sigma 0}}
   		-
   		T^2\, \tensor[_{\scriptscriptstyle{T}}]{c}{}^2\,
   		\tensor[_q]{c}{_{\alpha}}
   	\right)\\\notag
   	& \cdot\, \left(
   		\tensor{q}{_{\beta}}
   		+
   		\dot{T}\, \tensor[_{\scriptscriptstyle{T}}]{c}{}\,
   		\tensor{\hat{C}}{^{\gamma}_{\sigma \beta}}\, \tensor{\zeta}{_{\gamma 0}}\,
   		\tensor{\zeta}{^{\sigma 0}}
   		-
   		T^2\, \tensor[_{\scriptscriptstyle{T}}]{c}{}^2\,
   		\tensor[_q]{c}{_{\beta}}
   	\right)\\\notag
   	&-
   	\tensor{h}{_{\alpha \beta}}
   	\left(
   		\frac{1}{3}\, \rho
   		+
   		\left(
   			\frac{\dot{T}}{T}
   		\right)^2
   		\tensor{\zeta}{^{0 0}}
   		-
   		\frac{1}{3}\, T^2\, \tensor[_{\scriptscriptstyle{T}}]{c}{}^2\,
   		\tensor[_{\rho}]{c}{}
   	\right)\\
   	&+
   	2\, \dot{T}\, \tensor[_{\scriptscriptstyle{T}}]{c}{}\,
   	\tensor{\hat{C}}{^{\rho}_{\kappa (\alpha}}\, \tensor{\zeta}{_{\beta) \rho}}\,
   	\tensor{\zeta}{^{\kappa 0}}
   	+
   	T^2\, \tensor[_{\scriptscriptstyle{T}}]{c}{}^2\,
   	\tensor[_{\pi}]{c}{_{\alpha \beta}} \, . \label{eq:GUSTAV2}
\end{align}
As a constitutive quantity
$\tensor{\pi}{_{\alpha \beta}}$ is a function which is
linear and quadratic in the heat-flow and linear
in the energy density, while the temperature is also
included with its first and second time derivative.

The consideration of simple models like non-tilted or
stationary ones leads back to, e.g., the well-known Friedmann
or the G\"odel space-times (in both cases the constants of integration
are determined; see Sec. \ref{sec:matter-eqs}) and similar anisotropic models as
discussed in \cite{coley1994}.
In this context the expressions \eqref{eq:energy_dens_sol_final2} and
\eqref{eq:druck_sol_final2} or \eqref{eq:energy_dens_sol_final-re} and
\eqref{eq:druck_sol_final-re}, respectively, can be understood as generalized
Friedmann equations.

By rewriting the equations \eqref{eq:energy_dens_sol_final2},
\eqref{eq:druck_sol_final2}, \eqref{eq:heatflow_propag_sol_final2} and
\eqref{eq:anisotrop_druck_sol_final2_test} in terms of the observational
quantities $H=\frac{\dot{a}}{a}$ for the Hubble function and
$\left( \frac{\dot{a}}{a} \right)\frac{\dot{}}{}=-H^2(1+q)$ for the deceleration
parameter $q$ one receives limits on acceleration, rotation, 
heat-flux and anisotropic pressure.

The corresponding equations take the form
\begin{align} \label{eq:HEINZ}
%energy dens.
    \rho
   ={}&
    \tensor{\zeta}{^{0 0}}
    H^2 \left( 2q-1 \right)
    +
    2\, H^2\, \left( 1+q \right)
    -
    2\, H\, \frac{1}{a}\, \tensor{\hat{C}}{^{\gamma}_{\kappa \gamma}}\, \tensor{\zeta}{^{\kappa 0}}
    +
    \frac{1}{a^2}\, \tensor[_\rho]{c}{}\,,\\
  \label{eq:PETER}
%pressure
    p
  ={}&
    \frac{1}{3}\, \tensor{\zeta}{^{00}}\, H^2  \left( 5-4\, q \right)
    +
    \frac{2}{3}\, H^2\, \left( 1+q \right)
    +
    \frac{4}{3}\, \frac{H}{a}\, \tensor{\hat{C}}{^{\gamma}_{\kappa \gamma}}\, \tensor{\zeta}{^{\kappa 0}}
    +
    \frac{1}{a^2}\, \tensor[_p]{c}{}\,,\\
%heat flux
    \tensor{q}{_{\alpha}}
  ={}&
    - 2\, H^2\, \left( 1+q \right)\, \tensor*{h}{*^0 _{\alpha}}
    +
    \frac{H}{a}\,
    \tensor{\hat{C}}{^{\gamma} _{\kappa\alpha}}\, \tensor{\zeta}{_{\gamma 0}}\, \tensor{\zeta}{^{\kappa 0}}
    +
    \frac{1}{a^2}\, \tensor[_q]{c}{_{\alpha}}\,,
    \\ \notag
%anisotrop. press.
    \tensor{\pi}{_{\alpha \beta}}
  ={}&
    - \frac{2}{3}\, H^2\, \left( 1+q \right)\,
    \left(
      \tensor{h}{_{\alpha \beta}}
      \left(
	\tensor{\zeta}{^{0 0}} + 1
      \right)
      -
      3\, \tensor*{h}{*_{\alpha} ^0}\, \tensor*{h}{*_{\beta} ^0}
    \right)\\ \label{eq:HEINZ_last}
    &+
    \frac{2}{3}\, \frac{H}{a}\, \tensor{\hat{C}}{^{\rho}_{\kappa \mu}}\, \tensor{\zeta}{^{\kappa 0}}\,
    \left(
      \tensor*{\delta}{*^{\mu}_{\rho}}\, \tensor{h}{_{\alpha \beta}}
      -
      3\, \tensor*{\delta}{*^{\mu}_{(\alpha}}\, \tensor{\zeta}{_{\beta)\rho}}
    \right)
    +
    \frac{1}{ a^2}\, \tensor[_\pi]{c}{_{\alpha \beta}}\,,
\end{align}
so that the matter content can be described by the observable quantities $H$,
$q$ and the model-dependent constants $\zeta^{00}$, $\tensor*{h}{^0_{\alpha}}$, $\hat{C}^{\alpha}{}_{\beta\gamma}$
as well as the constants
of integration, eventually given by initial or boundary conditions.

In analogy to the calculations which lead to \eqref{eq:HEINZ2},
one obtains from \eqref{eq:HEINZ} and \eqref{eq:PETER}
\begin{align} \label{eq:HUGO}
    2\, \rho + p
    	=
    	3\, H^2\,
    	\left(
    		2\, q + 2 + \tensor{\zeta}{^{0 0}}
    	\right)
    	+
    	\frac{1}{a^2}\,
    	\left(
    		 2\, \tensor[_{\rho}]{c}{}
    		 +
    		 3\, \tensor[_{p}]{c}{}
    	\right)
\end{align}
which can again be regarded as an equation of state given by observational quantities. The class of models we consider here may have an
anisotropic behavior of the Hubble flow and the galaxy distribution
function \cite{obukhov1990bianchi_list,obukhov2000}. In this context, the observation of a large-scale flow of galaxies,
called ``dark flow'', with respect to the CMB is remarkable (see
\cite{kashlinsky2012} for a review). A detailed discussion of this and other possibly observable effects in non-rotating models can be found in \cite{CLARK1, CLARK2, CLARK3}.

Refraining from possible further restrictions on the relations \eqref{eq:HEINZ}
- \eqref{eq:HEINZ_last}, one finds the following hypothetical scenario.

For a large scale factor $a\gg 0$, the structure constants and the constants of integration are negligible, such that, 
for the behavior of the matter content, the expansion rate $H$ and the deceleration rate $q$ are most important. 
Moreover, one sees that $q$ has critical values at which the behavior of the
matter variables changes. For instance in the case of large accelerations ($q<-1$), which for
cosmological models  means a strongly increasing expansion and for local models
(like stars) a strongly increasing collapse, most matter variables change the
sign.
All matter variables display generally the same dependence on the expansion rate $H$ and are therefore of likewise importance. 
A more detailed discussion can only be achieved if the dependence on
$\zeta^{00}$ and $\tensor*{h}{*^0_{\alpha}}$ is fixed for specified
Bianchi models.

For small values of the scale factor $a$, i.e. in the early cosmological
phase or for objects which become very dense,
the structure and the integration constants
become much more important in comparison to $H$ and $q$.
Besides, all matter variables show the same behavior and are therefore of equal
importance.
When the scale factor $a(t)$ increases, the heat-flux and the
anisotropic pressure essentially behave like the energy and the pressure,
they dilute.

\appendix
%\section{Appendix} 
\label{sec:appendix}
\section{Kinematic relations} \label{sec:apendix_kin-rels}
The following relations between the kinematic invariants are used to
obtain and simplify results of the Sections \ref{sec:matter-eqs} and
\ref{sec:Interpretation} in kinematic terms.
\begin{align} \label{eq:acc_div}
		\tensor{\dot{u}}{^{\gamma}_{\Arrowvert \gamma} }
	={}&
		\frac{1}{3}
		\left(
		    \left(
		      \dot{\Theta} + \Theta^2
		    \right)
		    \left( \tensor{\zeta}{^{0 0}} + 1 \right)
		    +
		    T\, \Theta\, \tensor{\hat{C}}{^{\gamma}_{\kappa \gamma}}\, \tensor{\zeta}{^{\kappa 0}}
		\right)\, ,\\
%\end{align}
%
%\begin{align} 
\label{eq:acc_squ}
		\dot{u}^2
	={}&
		\frac{1}{9}\, \Theta^2\, \left( \tensor{\zeta}{^{0 0}} + 1 \right)
	=
		\frac{1}{9}\, \Theta^2\, \tensor{h}{^{0 0}}\, ,\\
%\end{align}
%
%\begin{align} 
	\notag \label{eq:grad_acc}
    \tensor{\dot{u}}{_{\alpha \Arrowvert \beta}}
  ={}&
    \frac{1}{3}\, \dot{\Theta}\, \tensor*{h}{*_{\alpha}^{0}}\, \tensor*{\delta}{*_{\beta}^{0}}
    +
    \dot{u}^2
    \left(
      \tensor{h}{_{\alpha \beta}}
      +
      \tensor{u}{_{\alpha}}\, \tensor{u}{_{\beta}}
    \right)
    +
    \frac{1}{3}\, \Theta\,
    \left(
      \tensor{u}{_{\alpha}}\, \tensor{\dot{u}}{_{\beta}}
      +
      \tensor{\omega}{_{\alpha \beta}}
    \right) \\
    &-
    \tensor{\dot{u}}{_{\alpha}}\, \tensor{\dot{u}}{_{\beta}}
    -
    \tensor{\omega}{_{\kappa (\alpha}}\, \tensor{u}{_{\beta)}}\, \tensor{\dot{u}}{^{\kappa}}
    -
    \frac{1}{2}\, T\, \tensor{\dot{u}}{^{\kappa}}\,
    \tensor{\hat{C}}{^{\mu}_{\kappa (\alpha}}\, \tensor{h}{_{\beta) \mu}}\,,\\
%\end{align}
%
%and
%
%\begin{align} 
\label{eq:div_rot}
		\tensor{\omega}{^{\tau \mu}_{\Arrowvert \mu}}\,
		\tensor{h}{_{\alpha \tau}}
	={}& \tensor{\omega}{_{\alpha \mu}}\, \tensor{\dot{u}}{^{\mu}}
		- \frac{1}{2}\, T\, \tensor{\omega}{_{\nu \kappa}}\,
		\tensor{\hat{C}}{^{\tau}_{\mu \sigma}}\, \tensor{\zeta}{^{\mu \kappa}}\,
		\tensor{\zeta}{^{\nu \sigma}}\, \tensor{h}{_{\alpha \tau}}
		+
		T\, \tensor{\omega}{_{\alpha \kappa}}\,
		\tensor{\hat{C}}{^{\rho}_{\mu \rho}}\, \tensor{\zeta}{^{\mu \kappa}} \, .
\end{align}
\section{Tetrad formulation of curvature} \label{sec:appendix-curv}
The connection coefficients in the tetrad formulation (see Sec.
\ref{sec:tetrads-kin-invariants}), the so-called Ricci rotation coefficients, can be expressed in terms of the Christoffel symbols
$\tensor{\Gamma}{_c^d_b}$,
\begin{align} \label{eq:konnektion_bez_christoffel-riccirotkoeff}
	\tensor{\Omega}{_c^\mu_\nu}
	=
	\tensor{\theta}{^\mu_d}\, \tensor{\theta}{_\nu^b}\, \tensor{\Gamma}{_c^d_b}
	-
	\tensor{\theta}{_\nu^b}\, \tensor{\theta}{^\mu_{b,c}} \, ,
\end{align}
or, due to \eqref{eq:metric_alter_tetrad} by the structure constants
\eqref{eq:metric_struct-const_embed}, respectively,
\begin{align} \notag
  \tensor{\Omega}{_{\rho}^{\mu}_{\nu}}
  ={}&
  \frac{\dot{a}}{a}\,
  \left(
    \tensor*{\delta}{*_{\rho}^{\mu}}\, \tensor*{\delta}{*_{\nu}^{0}}
    -
    \tensor{\zeta}{_{\nu \rho}}\, \tensor{\zeta}{^{\mu 0}}
  \right) \\
  {}&+
  \frac{1}{2\, a}\,
  \tensor{\zeta}{^{\kappa \mu}}\,
  \left(
    \tensor{\hat{C}}{^{\gamma}_{\nu \rho}}\, \tensor{\zeta}{_{\kappa \gamma}}
    + \tensor{\hat{C}}{^{\gamma}_{\rho \kappa}}\, \tensor{\zeta}{_{\nu \gamma}}
    - \tensor{\hat{C}}{^{\gamma}_{\kappa \nu}}\, \tensor{\zeta}{_{\rho \gamma}}
  \right) \, .
\end{align}
Determining the Riemannian curvature tensor by the Ricci-identity and the
tetrads,
\begin{align}
  \tensor{R}{_{m b c d}}\, \tensor{\theta}{_{\alpha}^{m}}
  =
  \tensor{\theta}{_{\alpha b; c; d}} - \tensor{\theta}{_{\alpha b; d; c}} \, ,
\end{align}
the Ricci tensor can be brought to the form
\begin{align} \label{eq:ricci_tensor_struct-const} \notag
	\tensor{R}{_{\alpha \beta}}
	={}&
	- \left( \frac{\dot{a}}{a} \right)_{,0}
	\left(
		2\, \tensor*{\delta}{*_{\alpha}^0}\, \tensor*{\delta}{*_{\beta}^0}
		+
		\tensor{\zeta}{^{0 0}}\, \tensor{\zeta}{_{\alpha \beta}}
	\right)
	-
	3\, \left( \frac{\dot{a}}{a} \right)^2 \, \tensor{\zeta}{^{0 0}}\, \tensor{\zeta}{_{\alpha \beta}}
	\\
	&-
	\frac{\dot{a}}{a^2}
	\left(
		\tensor{\hat{C}}{^{\gamma}_{\kappa \beta}}\, \tensor{\zeta}{^{\kappa 0}}\,
		\tensor{\zeta}{_{\alpha \gamma}}
		+
		\tensor{\hat{C}}{^{\gamma}_{\kappa \alpha}}\, \tensor{\zeta}{^{\kappa 0}}\,
		\tensor{\zeta}{_{\beta \gamma}}
		+
		\tensor{\hat{C}}{^{\rho}_{\mu \rho}}\, \tensor{\zeta}{^{\mu 0}}\,
		\tensor{\zeta}{_{\alpha \beta}}
	\right)
	-
	\frac{1}{2\, a^2}\, \tensor{\tilde{R}}{_{\alpha \beta}} \,
\end{align}
with
\begin{align} \label{eq:ricci_tensor_3dim} \notag
	\tensor{\tilde{R}}{_{\alpha \beta}}
	={}&
	- \frac{1}{2}
	\left(
		- \tensor{\hat{C}}{^{\gamma}_{\kappa \beta}}\, \tensor{\hat{C}}{^{\rho}_{\mu
		\rho}}\, \tensor{\zeta}{_{\alpha \gamma}}\, \tensor{\zeta}{^{\mu \kappa}}
		+
		\tensor{\hat{C}}{^{\mu}_{\nu \beta}}\, \tensor{\hat{C}}{^{\nu}_{\alpha
		\mu}}
		\right.\\\notag
		& \left.\vphantom{}
		-
		\tensor{\hat{C}}{^{\gamma}_{\kappa \alpha}}\, \tensor{\hat{C}}{^{\rho}_{\mu
		\rho}}\, \tensor{\zeta}{_{\beta \gamma}}\, \tensor{\zeta}{^{\mu \kappa}}
		-
		\tensor{\hat{C}}{^{\mu}_{\nu \beta}}\, \tensor{\hat{C}}{^{\tau}_{\sigma
		\alpha}}\, \tensor{\zeta}{_{\mu \tau}}\, \tensor{\zeta}{^{\nu \sigma}}
		\right.\\
		& \left.\vphantom{}
		+
		\frac{1}{2}\,
		\tensor{\hat{C}}{^{\gamma}_{\kappa \nu}}\, \tensor{\hat{C}}{^{\tau}_{\mu
		\sigma}}\, \tensor{\zeta}{_{\beta \gamma}}\, \tensor{\zeta}{^{\mu \kappa}}\,
		\tensor{\zeta}{_{\alpha \tau}}\, \tensor{\zeta}{^{\nu \sigma}}
	\right)\, .
\end{align}
Accordingly, the Ricci scalar becomes
\begin{align} \label{eq:ricci_scalar_struct-const} %\notag
      R ={}&
	- 6 \left(
			\frac{\dot{a}}{a}
		\right)_{,0}
	\tensor{\zeta}{^{0 0}}
	-
	12\, \left( \frac{\dot{a}}{a} \right)^2 \tensor{\zeta}{^{0 0}}
	-
	6\, \frac{\dot{a}}{a^2}\, \tensor{\hat{C}}{^\rho _{\mu \rho}}\,
	\tensor{\zeta}{^{\mu 0}}
	-
	\frac{1}{a^2}\, \tilde{R}  \,
\end{align}
with
\begin{align} \label{eq:ricci_scalar_3dim} %\notag
	\tilde{R}
	={}&
	\tensor{\hat{C}}{^{\gamma}_{\kappa \gamma}}\,
	\tensor{\hat{C}}{^{\rho}_{\mu \rho}}\, \tensor{\zeta}{^{\mu \kappa}}
	- \frac{1}{2}\, \tensor{\hat{C}}{^{\mu}_{\nu \beta}}\,
	\tensor{\hat{C}}{^{\nu}_{\alpha \mu}}\, \tensor{\zeta}{^{\alpha \beta}}
	+ \frac{1}{4}\, \tensor{\hat{C}}{^{\mu}_{\nu \beta}}\,
	\tensor{\hat{C}}{^{\tau}_{\sigma \alpha}}\, \tensor{\zeta}{^{\nu \sigma}}\,
	\tensor{\zeta}{_{\mu \tau}}\, \tensor{\zeta}{^{\alpha \beta}} \, .
\end{align}
The expressesions
$\tensor{\tilde{R}}{_{\hat{\alpha} \hat{\beta}}}$ of
 \eqref{eq:ricci_tensor_3dim} and equivalently $\tilde{R}$
 of \eqref{eq:ricci_scalar_3dim}
 can be identified with the Ricci tensor and the Ricci scalar
 of 3-dimensional Bianchi spaces \cite{wald1984}.

This results in the following shape of the Einstein tensor:
\begin{align} \notag
	\tensor{G}{_{\alpha \beta}}
	={}&
	\tensor{R}{_{\alpha \beta}} - \frac{1}{2}\, R\, \tensor{\zeta}{_{\alpha
	\beta}}\\\notag ={}&
	2 \left(\frac{\ddot{a} -\dot{a}^2}{a^2}\right)
	\left(
		\tensor{\zeta}{_{\alpha \beta}}\, \tensor{\zeta}{^{0 0}}
		- \tensor*{\delta}{*_{\alpha}^{0}}\, \tensor*{\delta}{*_{\beta}^{0}}
	\right)
	+
	3 \left(
		\frac{\dot{a}}{a}
	\right)^2
	\tensor{\zeta}{_{\alpha \beta}}\,
	\tensor{\zeta}{^{0 0}}\\
	&+
	2 \frac{\dot{a}}{a^2}\,
	\tensor{\hat{C}}{^{\rho}_{\kappa \mu}}\, \tensor{\zeta}{^{\kappa 0}}
	\left(
		\tensor*{\delta}{*_{\rho}^{\mu}}\, \tensor{\zeta}{_{\alpha \beta}}
		- \tensor*{\delta}{*_{(\alpha}^{\mu}}\, \tensor{\zeta}{_{\beta) \rho}}
	\right)
	+
	\frac{1}{a^2}
	\left(
		- \tensor{\tilde{R}}{_{\alpha \beta}}
		+ \frac{1}{2}\tilde{R}\, \tensor{\zeta}{_{\alpha \beta}}
	\right) \, .
\end{align}

\section{Constants of integration} \label{sec:appendix-const-of-int}
From the
field equations, $\tensor{G}{_{\alpha \beta}} = \tensor{T}{_{\alpha \beta}}$,
in terms of the tetrad formulation from Sec.
\ref{sec:tetrads-kin-invariants} and together with \eqref{eq:metric_allg_obuk},
one finds
\begin{align}
      - \tensor{\tilde{R}}{_{\alpha \beta}} + \frac{1}{2}\, \tilde{R}\,
      \tensor{\zeta}{_{\alpha \beta}} =
      \tensor[_{\scriptscriptstyle{EI}}]{c}{_{\alpha \beta}} \, .
\end{align}
Then, because of \eqref{eq:int_const_general_ei} the constants of integration, i.e. $\tensor[_{\rho}]{c}{}$, $\tensor[_{p}]{c}{}$,
$\tensor[_q]{c}{_{\alpha}}$ and $\tensor[_\pi]{c}{_{\alpha \beta}}$,
for the energy density, the isotropic pressure, the heat-flux and the
anisotropic pressure \eqref{eq:energy_dens_sol_final2} -
\eqref{eq:anisotrop_druck_sol_final2} become, in this order,
\begin{align} %\notag 
\label{eq:int_const_energy-dens}
      \tensor[_\rho]{c}{}
    ={}&
      \tensor[_{\scriptscriptstyle{EI}}]{c}{_{\alpha \beta}}\,
      \tensor{u}{^{\alpha}}\, \tensor{u}{^{\beta}} %\\
    =%{}&
      - \frac{1}{2}\, \tilde{R} + 2\, \left(
      \frac{\omega}{\tensor[_{\scriptscriptstyle{T}}]{c}{}\, T} \right)^2 \, ,\\
%\end{align}
%
%\begin{align} %\notag 
\label{eq:int_const_druck}
      \tensor[_p]{c}{}
    ={}&
      \frac{1}{3}\, \tensor[_{\scriptscriptstyle{EI}}]{c}{_{\alpha \beta}}\,
      \tensor{h}{^{\alpha \beta}} %\\
    =%{}&
      \frac{1}{6}\, \tilde{R} + \frac{2}{3}\, \left(
      \frac{\omega}{\tensor[_{\scriptscriptstyle{T}}]{c}{}\, T} \right)^2\, ,\\
%\end{align}
%
%\begin{align} \notag 
\label{eq:int_const_heat-flow}
      \tensor[_q]{c}{_{\alpha}}
    ={}&
      - \tensor[_{\scriptscriptstyle{EI}}]{c}{_{\beta \gamma}}\,
      \tensor{u}{^\beta}\, \tensor*{h}{*_{\alpha}^{\gamma}} %\\
    =%{}&
      \frac{1}{\tensor[_{\scriptscriptstyle{T}}]{c}{}^2\, T^2}
	\left(
		\tensor{\omega}{^{\kappa \mu}_{\Arrowvert \mu}}\,
		\tensor{h}{_{\alpha \kappa}}
		-
		\tensor{\omega}{_{\alpha \mu}}\, \tensor{\dot{u}}{^{\mu}}
	\right) \,,\\
%\end{align}
%
%and
%
%\begin{align} 
\notag \label{eq:int_const_anisotrop-druck}
      \tensor[_\pi]{c}{_{\alpha \beta}}
    ={}&
      \tensor[_{\scriptscriptstyle{EI}}]{c}{_{\gamma \delta}}\,
      \tensor*{h}{*_{\alpha}^{\gamma}}\, \tensor*{h}{*_{\beta}^{\delta}}
		- \tensor[_p]{c}{}\, \tensor{h}{_{\alpha \beta}} \\\notag
    ={}&
	- \tensor{\tilde{R}}{_{\alpha \beta}}
	+
	\frac{1}{3}\, \tilde{R}\, \tensor{h}{_{\alpha \beta}}
	-
	\frac{2}{3}\, \left( \frac{\omega}{\tensor[_{\scriptscriptstyle{T}}]{c}{}\, T}
	\right)^2 \left( 3\, \tensor{u}{_{\alpha}}\, \tensor{u}{_{\beta}}
	  +
	  \tensor{h}{_{\alpha \beta}}
	\right)\\
	&-
	\frac{2}{\tensor[_{\scriptscriptstyle{T}}]{c}{}^2\, T^2}
	\left(
	  \tensor{\omega}{_{\rho (\alpha}}\, \tensor{u}{_{\beta)}}\, \tensor{\dot{u}}{^{\rho}}
	  +
	  \tensor{\omega}{^{\tau \mu}_{\Arrowvert \mu}}\, \tensor{h}{_{\tau (\alpha}}\, \tensor{u}{_{\beta)}}
	\right)\, ,
\end{align}
where in \eqref{eq:int_const_heat-flow} and \eqref{eq:int_const_anisotrop-druck}
it was made use of the relation \eqref{eq:div_rot}.
If one reinserts the constants of integration
\eqref{eq:int_const_energy-dens} - \eqref{eq:int_const_anisotrop-druck}
into the matter equations
\eqref{eq:energy_dens_sol_final2} -
\eqref{eq:anisotrop_druck_sol_final2}, they take the purely kinematic forms
\eqref{eq:energy_dens_sol_final-re} - \eqref{eq:anisotrop_druck_sol_final2-ext}.

The \textit{summarized} constants of integration, $\tensor[_\rho]{c}{} $,
$\tensor[_p]{c}{}$, $\tensor[_q]{c}{_\alpha}$ and $\tensor[_\pi]{c}{_{\alpha
\beta}}$, in this paper
are pieced together as follows:
\begin{align} \label{eq:energy_dens_int-const}
	\tensor[_\rho]{c}{}
	:={}& -\tensor[_\rho]{\tilde{c}}{}
		- 2\, c_1\,
			\left(
	 			2\, \tensor{\zeta}{^{0 0}} - 1
	 		\right)
		- 2\, c_2\, \left(\tensor{\zeta}{^{0 0}} +1 \right)
		- 2\, c_3\, \tensor{\hat{C}}{^\gamma_{\kappa \gamma}}\,
	\tensor{\zeta}{^{\kappa 0}}\, ,\\
%\end{align}
%
%\begin{align} 
\label{eq:druck_int-const}
	\tensor[_p]{c}{}
	:={}&
	- \tensor[_p]{\tilde{c}}{}
	- \frac{2}{3}\, c_2 \left( 1 - 2\, \tensor{\zeta}{^{0 0}} \right)
	+ \frac{2}{3}\, c_1\,
	\left(
		7\, \tensor{\zeta}{^{0 0}} + 1
	\right)
	+ \frac{4}{3}\, c_3\, \tensor{\hat{C}}{^\gamma_{\kappa \gamma}}\,
	\tensor{\zeta}{^{\kappa 0}}\, ,\\
%\end{align}
%
%\begin{align}
  \tensor[_q]{c}{_\alpha}
  :={}&
    - \tensor[_q]{\tilde{c}}{_\gamma}
    \left(
		\tensor*{h}{*_{\alpha}^{\gamma}}
		- \tensor*{h}{*_{\alpha}^{0}}\, \tensor{u}{^{\gamma}}
	\right)
    -
    2\, \left(c_1 + c_2 \right)\, \tensor*{h}{*_{\alpha}^0}
    - c_3\, \tensor{\hat{C}}{^{\gamma}_{\alpha \sigma}}\,
    \tensor{\zeta}{_{\gamma 0}}\,
    \tensor{\zeta}{^{\sigma 0}}\, ,%\\
\end{align}
\begin{align} 
\notag
	\tensor[_\pi]{c}{_{\alpha \beta}}
	:={}&
	\frac{1}{3}
	\left(
	      - \tensor[_\pi]{\tilde{c}}{_{\gamma \delta}}
	 		\left(
	 			\tensor*{h}{*_{\alpha}^{\gamma}}
	 			- \tensor*{h}{*_{\alpha}^{0}}\, \tensor{u}{^{\gamma}}
	 		\right)
	 		\left(
	 			\tensor*{h}{*_{\beta}^{\delta}}
	 			- \tensor*{h}{*_{\beta}^{0}}\, \tensor{u}{^{\delta}}
	 		\right)
	 		\vphantom{\frac{}{}}\right.\\ \notag
			& \left.\vphantom{\frac{}{}}
	      -
	      2\, \left(
		c_1 - c_2
	      \right)
	      \left(
		      \tensor{h}{_{\alpha \beta}}
		      \left(  \tensor{\zeta}{^{0 0}} + 1  \right)
		      -
		      3\, \tensor*{h}{*_{\alpha}^0}\, \tensor*{h}{*_{\beta}^0}		
	      \right)
	      \right.	\\
	{}& \left.\vphantom{\frac{}{}}
	      +
	      2\, c_3 \,
	      \tensor{\hat{C}}{^\rho_{\mu \kappa}}\, \tensor{\zeta}{^{\kappa 0}}
	      \left(
		      \tensor*{\delta}{*_{\rho}^\mu}\, \tensor{h}{_{\alpha \beta}}
		      -
		      3\, \tensor*{\delta}{*_{(\alpha}^\mu}\, \tensor{h}{_{\beta) \rho }}		
	      \right)						
	\right)\, .
\end{align}
The occurring objects $c_1$, $c_2$, $c_3$, $\tensor[_\rho]{\tilde{c}}{}$,
$\tensor[_p]{\tilde{c}}{}$, $\tensor[_q]{\tilde{c}}{_{\hat{\alpha}}}$ and
$\tensor[_\pi]{\tilde{c}}{_{\hat{\alpha} \hat{\beta}}}$ are the \textit{actual}
constants of integration yielded by the following integrals, which are to be
calculated in Subsec. \ref{sec:matter-eqs_non-perfect-fluid}:
\begin{gather} \label{eq:energy_dens_int-const1_part1}
		\int
   		\ddot{a}\, \dot{a}\,
		\mathrm{d}x^0
	= \frac{\dot{a}^2}{2} + c_1\, , \;
		\int
   		\dddot{a}\, a\,
		\mathrm{d}x^0
	= \ddot{a}\, a
		- \frac{\dot{a}^2}{2} + c_2 \, , \;
	\int
   		\ddot{a}\,
		\mathrm{d}x^0
	= \dot{a} + c_3 \, ,\\
%\end{align}
%
%\begin{align}  
\label{eq:energy_dens_int-const1}
	 \int
    \left(a^2\, \rho \right)_{,0}  \mathrm{d}x^0
	 = a^2\, \rho
	 		+ \tensor[_\rho]{\tilde{c}}{} \, , \quad
	 \int
    \left(a^2\, p \right)_{,0}  \mathrm{d}x^0
	 = a^2\, p
	 		+ \tensor[_p]{\tilde{c}}{} \, ,\\
%\end{gather}
%
%\begin{align}  
\label{eq:heat-flow_int-const1}
	 \int
    \left(a^2\, \tensor{q}{_\alpha} \right)_{,0}  \mathrm{d}x^0
	 = a^2\, \tensor{q}{_\alpha}
	 		+ \tensor[_q]{\tilde{c}}{_\beta}
	 		\left(
	 			\tensor*{h}{*_{\alpha}^{\gamma}}
	 			- \tensor*{h}{*_{\alpha}^{0}}\, \tensor{u}{^{\gamma}}
	 		\right)\, ,\\
%\end{gather}
%
%\begin{align}  
\label{eq:anisotrop-druck_int-const1}
	 %&
	 \int
    3\, \left(a^2\, \tensor{\pi}{_{\alpha \beta}} \right)_{,0}
    \mathrm{d}x^0
     = a^2\, \tensor{\pi}{_{\alpha \beta}}
	 		+ \tensor[_\pi]{\tilde{c}}{_{\gamma \delta}}
	 		\left(
	 			\tensor*{h}{*_{\alpha}^{\gamma}}
	 			- \tensor*{h}{*_{\alpha}^{0}}\, \tensor{u}{^{\gamma}}
	 		\right)
	 		\left(
	 			\tensor*{h}{*_{\beta}^{\delta}}
	 			- \tensor*{h}{*_{\beta}^{0}}\, \tensor{u}{^{\delta}}
	 		\right) \, .
\end{gather}
%
%
%
%\nocite{*} %alle Literaturhinweise anzeigen!
%\bibliographystyle{spphys}
%\bibliography{references}

\end{document}